\documentclass[aps,prresearch,reprint,floatfix]{revtex4-2}
\usepackage{graphicx,xcolor,amssymb,amsmath,hyperref,dcolumn,bm}
\usepackage[print-unity-mantissa=false]{siunitx}
\PassOptionsToPackage{normalem}{ulem}%%%%%%%
\usepackage{ulem}
\usepackage{physics}
\usepackage[version=4]{mhchem}

\graphicspath{{Figures/}}

\begin{document}

\title{Active particle in a very thin interfacial droplet}
\author{Airi N. Kato$^{1}$}%
%\email{airi.nakamoto@u-bordeaux.fr}
\author{Kaili Xie$^{1,2}$}
\author{Benjamin Gorin$^{1}$}
\author{Jean-Michel Rampnoux$^{1}$}
%\author{Alois W\"{u}rger$^{1}$}
\author{Hamid Kellay$^{1}$}%
 \email{hamid.kellay@u-bordeaux.fr}
\affiliation{
$^{1}$ Laboratoire Ondes et Mati\`ere d’Aquitaine, Universit\'e de Bordeaux, Talence 33405, France
}%
\affiliation{%
 $^{2}$Van der Waals-Zeeman Institute, Institute of Physics, University of Amsterdam, 1098XH Amsterdam, The Netherlands
}%

\date{\today}
%https://journals.aps.org/prl/info/infoL.html
%fig1 AR=1.8 single
%fig2 AR=1.2 single
%fig3 AR=3.7 double
%fig4 AR=4.0 double
%[(150 / aspect ratio) + 20 words] for single-column figures, and [300 / (0.5 * aspect ratio)] + 40 words for double-column figures.
% in total 640 words
%3,750 words
%2915+16*8+640=3683
\begin{abstract}%within 600characters including space for PRL!
A single light-driven Janus particle confined in a very thin oil droplet at an air--water interface displays intriguing dynamics. While laser activation induces rapid horizontal motion (\SI{1}{\milli\metre \per\second}--\SI{1}{\centi\metre\per\second}) by thermal Marangoni flow, the particle exhibits unexpected periodic circular motions or intermittent irregular motions. We show that the trajectories are the result of a coupling between the self-propulsion of the particle and the spatiotemporal droplet thickness changes. We propose a simple model where the properties of the active particle trajectories are governed by capillary forces and torques due to the confinement of the particle in the thin droplet.  
\end{abstract}

% \textcolor{red}{A single light-driven Janus particle confined in a very thin oil droplet at an air–water interface displays intriguing dynamics. Upon laser activation, rapid horizontal motion (\SI{1}{\milli\metre\per\second}–\SI{1}{\centi\metre\per\second}) arises from thermal Marangoni flow, and depending on the contact angle, the particle also exhibits unexpected periodic circular or intermittent irregular motions. We show that spatiotemporal variations in droplet thickness govern these dynamics via capillarity, as a characteristic feature of thin film systems. We propose a scenario for periodic orbiting motion similar to that reported by Dauchot {\it et al.} [Phys. Rev. Lett. 122, 068002 (2019)], but with explicit quantification of the centripetal force and torque from nonequilibrium experimental profiles. This work points to new directions for active particle–based soft matter systems, where interfacial properties and activity together enable novel functionalities.}

\maketitle
%\textit{Introduction.---}%
Soft boundaries, such as droplets~\cite{Sanchez2012,adkins_dynamics_2022,xie_activity_2022,Kokot2022,Sakamoto2022,Sakuta2023}, vesicles~\cite{le_nagard_encapsulated_2022,vutukuri_active_2020,takatori2020active,Gu2025}, flexible membranes~\cite{deblais_boundaries_2018,junot_active_2017,boudet2021collections,paoluzzi2016shape}, and related systems~\cite{Alexandre2024} can serve as soft confinements that active particles can deform and reshape. These soft confinements are of interest in different contexts, both at the fundamental level \cite{paoluzzi2016shape} and for practical uses such as enhanced oil recovery~\cite{Hickl2022}, biofilm formation~\cite{Hickl2022, Prasad2023}, targeted drug delivery~\cite{Luo2018} and soft robotics \cite{deblais_boundaries_2018,boudet2021collections}.
%\textcolor{red}{Ideas: 1) conferring motility to cell-sized vesicles 2)distorted vesicle morphologies could improve the design of artificial systems such as small-scale soft robots and synthetic cells 3) soft robotics in general}

These boundaries also change the behavior of active particle assemblies~\cite{nikola2016active}. For example, vesicles can undergo significant deformation and possible propulsion due to the internal reorganization of the confined active components~\cite{le_nagard_encapsulated_2022,vutukuri_active_2020,takatori2020active}, offering design principles for microscale soft robots and synthetic cells.  
Droplet systems are particularly intriguing because the interface can trap or alter particle dynamics~\cite{wang_enhanced_2015,deng2020motile,Gidituri2022,Feng2022,Deng2022,Prasad2023}, opening pathways for the development of novel active interfacial materials.
The existence of interfaces can also induce Marangoni flows driven by concentration~\cite{Nakata1997,Yabunaka2012,Michelin2013,Koyano2019,ender2021surfactant,boniface2021role} or temperature gradients~\cite{wurger_thermally_2014,dietrich_microscale_2020}, offering a significantly more efficient and faster propulsion mechanism compared to self-diffusiophoresis.

Most studies to date have focused on the behavior of active particles and their mobility in bulk fluids; a few studies have also addressed this issue at single interfaces. However, the physical mechanisms underlying the interplay between activity and interfacial effects, in particular interfacial deformation as may occur when strong confinement is present, remain poorly understood.
%confined in three-dimensional compartments, where low interfacial tension or surface energy can lead to deformation by local mechanical pressure~\cite{adkins_dynamics_2022}, or where the active particles maintain shape stability through activity~\cite{xie_activity_2022}. 
Thin films~\cite{Wu2000,Sokolov2009,Mathijssen2016,Sankararaman2009}, thin droplets on substrates~\cite{Trinschek2020}, and interfaces~\cite{ash2012wetting,aveyard1999size} provide simplified yet representative systems to explore these questions. Such geometries could help clarify the roles of interfacial tension, active forces, and strong confinement and deformation in shaping particle behavior.

In this Letter, we study the dynamic behavior of a light-driven Janus particle (JP) strongly confined in a very thin, lens-like oil droplet at an air--water interface. Our main observation is that Marangoni-driven fast JP motions exhibit diverse temporal features depending strongly on the droplet geometry confining the particle. We show that the underlying mechanisms governing the particle trajectories arise from the dynamic coupling between the JP motions and the droplet deformation, clarifying the role of capillary confinement in setting active particle trajectories and motions. 
%In this Letter, we study the role of thin film droplet confinement for an active Janus particle (JP) with the presence of interfaces nearby. A Marangoni flow drives the JP very fast, exhibiting diverse trajectories with different temporal features accompanying the droplet thickness profile changes. We find the capillarity due to the confinement is essential 

\begin{figure}[tb]
\includegraphics[bb= 0 0 1519 844, width=0.5\textwidth]{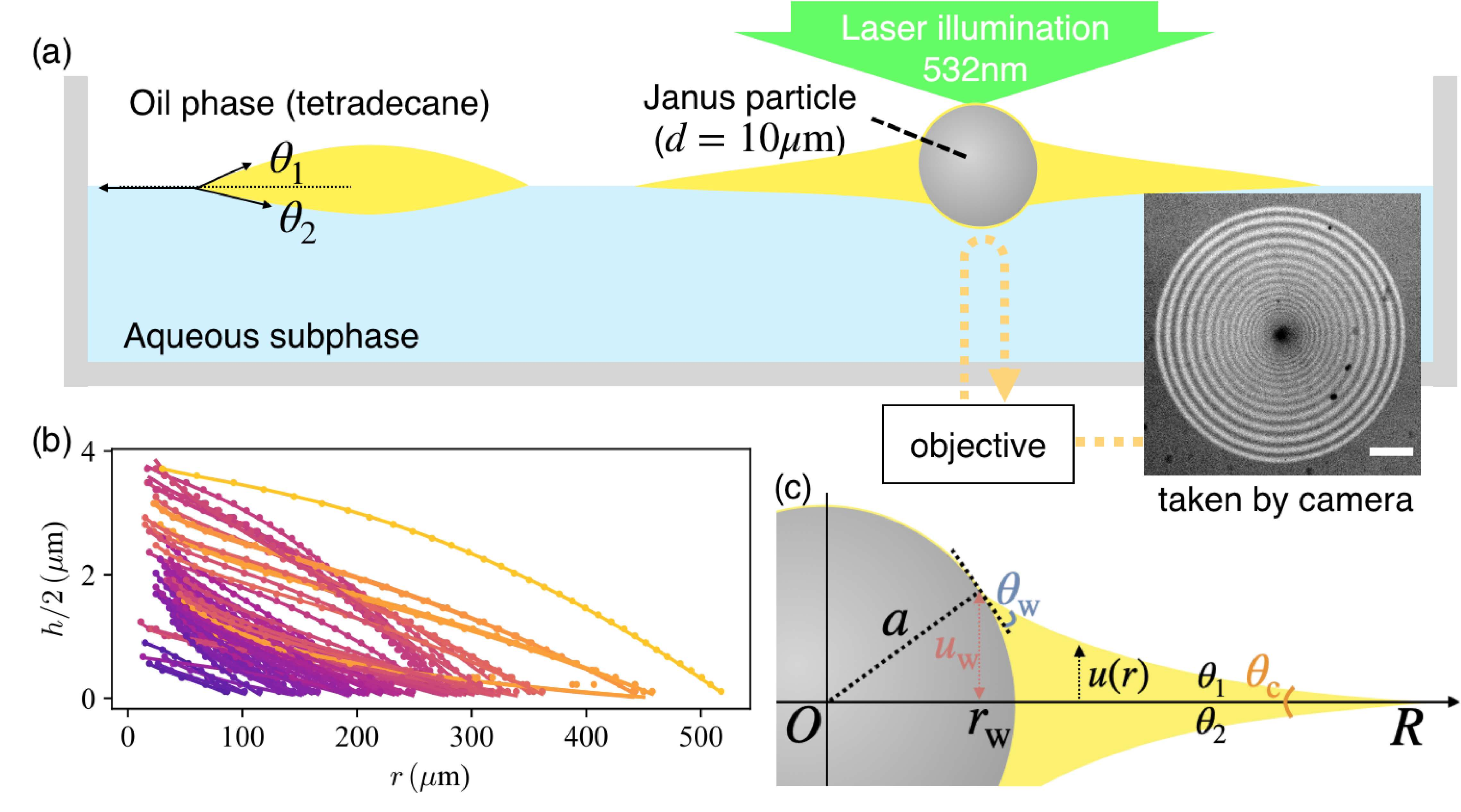}% Here is how to import EPS art
\caption{\label{fig:1} Experimental setup and thin oil droplet profiles with a Janus particle. (a) Schematics of the interfacial droplets with and without a JP. The inserted image is an example of a droplet with a JP (the black dot in the center). Scale bar=\SI{100}{\micro\metre}. (b) Interfacial profiles $u(r)=h(r)/2$ of droplets with a static JP. The dots correspond to the measured $(r_m, h_m/2)$ values using the fringes, and the solid lines are the fits to Eq.~\eqref{eq:u}. The colors are set by $R$. (c) Schematic of the lens-like geometry illustrating the contact angle $\theta_C$ and apparent contact angle $\theta_W$.} %(d) The probability distributions of the contact angles $\theta_\mathrm{c}$ and $\theta_\mathrm{w}$ before activation.}
\end{figure}

%\textit{Experimental method.---}%
The self-propelled Janus particles (JPs) used here are made of polystyrene spheres, of radius $a=\SI{5}{\micro\metre}$, with a half-coated metal layer (see details in Supplementary Material~\cite{supp_material}). These particles were suspended in tetradecane oil. 
The suspension was then placed onto an aqueous subphase (ultrapure water with 6.8 mM NaCl) in a Petri dish and chopped with a needle to create separated droplets with radii $R$ = \qtyrange{10}{600}{\micro\metre}, as illustrated in Fig.~\ref{fig:1}(a). We here focus primarily on the dynamics of a single JP within a droplet to avoid the complexity of multiple JPs such as aggregation in the thin droplet due to the long-range capillary bridge attraction~\cite{Willett2000}. The JP generally resides at the center of the droplet. 

To turn on the activity of the JP, a laser beam was used \cite{xie_activity_2022}. This activity can be tuned by the illumination intensity $I$ of the top-hat profile laser beam (wavelength: \SI{532} {\nano\metre} and radius: \SI{108} {\micro\metre})~\cite{xie_activity_2022}. Initially, the center of the illumination area was aligned with the center of the stationary droplet. The dynamics of the JP and the droplet were visualized using an inverted microscope (Axio Observer, Zeiss) equipped with a camera (Orca Flash 4.0 from Hamamatsu or Phantom v640 from Vision Research Inc.).

The static droplet radial thickness profiles $h(r)$ are shown in Fig.~\ref{fig:1}(b) and were obtained from the interference fringe pattern of the droplet as detailed below. The droplet is extremely thin and has a steep profile at the periphery of the JP as illustrated in the schematics of Fig.~\ref{fig:1}(a) and (c).

\smallskip

%\vspace{-2.5mm}
%\textit{Profile of the thin droplet.---}%
When the relationship $|\gamma_\mathrm{\it o}-\gamma_\mathrm{\it ow}|<\gamma_\mathrm{\it w}<\gamma_\mathrm{\it o}+\gamma_\mathrm{\it ow}$ holds for the interfacial tensions of $\gamma_\mathrm{\it o}$ (air--oil), $\gamma_\mathrm{\it w}$ (air--water) and $\gamma_\mathrm{\it ow}$ (oil--water), the particle-free oil droplets show a lens-like shape with small contact angles, $\theta_{1}\approx$ 0.022 rad and $\theta_{2}\approx$ 0.013 rad derived from the Young equation for the three fluids, $\gamma_w$ = $\gamma_o \cos{\theta_1}$ + $\gamma_{ow} \cos{\theta_2}$ (Fig.~\ref{fig:1}(a)).  
The particle-free droplets possess a quadratic profile for various radii (Fig.~S1(a)~\cite{supp_material}). In contrast, the droplet containing a JP forms a thin droplet with the JP centered, showing a circular symmetry confirmed by the fringes (see inset image of Fig.~\ref{fig:1}(a)). The radially averaged thickness profiles $h(r)$ can be reconstructed via fringe counting with the bright ring condition: $h_m=\frac{m\lambda}{2n_o}$ ($m=1,2,3,\cdots$. $\frac{\lambda}{2n_o}=225\SI{}{\nano\metre}$), where $\lambda$ and $n_o$ are the wavelength of the observation light (around $\SI{630}{\nano\metre}$) and the refractive index of oil ($n_o=1.43$). We assume that the most outward fringe corresponds to the phase difference $2\pi$, and the thickness changes monotonously and continuously. The series of bright spots and corresponding thicknesses $(r_m, h_m)$ give the thickness profile $h(r)$. 
From the Young--Laplace equation $\gamma \nabla^2 u(r)+p_{0}=0$, we can derive 
\begin{equation}\label{eq:u}
    u(r)=-u_0\ln\frac{r}{R}+u_2 \frac{R^2-r^2}{2R^2}
\end{equation}
which fitted well with all our measured interfacial profiles $(r_m, h_m/2)$. 
Here $r$ is the distance from the center of the drop, $u_0$ is a constant and $u_2=p_0R^2/\gamma$, where $p_{0}$ and $\gamma$ are the constant internal pressure and the interfacial tension. Here we assumed the symmetry of the oil droplet with $\theta_1=\theta_2$, the profile simplifies to $u(r)=h(r)/2$.
As defined in Fig.~\ref{fig:1}(c), $u_0,\,u_2,\,R$ can be determined from the droplet volume $V$ ~\cite{supp_material}, the three-phase contact angles $\theta_\mathrm{C}$, and the wetting angle $\theta_\mathrm{W}$. The angles $\theta_\mathrm{C}$ and $\theta_\mathrm{W}$ are estimated using the profiles in Fig.~\ref{fig:1}(b) as $\theta_\mathrm{C}\simeq 0.58^\circ \pm 0.30^\circ$ and $\theta_\mathrm{W}\simeq 23.52^\circ\pm 7.12^\circ$, where the ranges are the standard deviations. (see Fig.~S1~\cite{supp_material} for the probability distributions).

Since the angles $\theta_\mathrm{C}$ are material properties determined by the Young relations, the variation may originate from metastability by agitation when creating droplets. Note that $\theta_\mathrm{W}$ can be understood as the apparent angle because the JP is fully covered by the oil in this study. Therefore, no contact line pinning is expected. The JP has in-plane polarity in a random direction, unlike the interfacial trapped JPs, which have a propensity to have smaller in-plane polarity depending on the surface properties~\cite{Park2011,wang_enhanced_2015,Dietrich2017,dietrich2018active,dietrich_microscale_2020}. 
%The JP is stable at the center of the droplet without illumination. The static droplet radial thickness profiles $h(r)$ were reconstructed from the interference fringe pattern of the droplet by the method in Appendix A, and shown in Fig.~\ref{fig:1}(b). The droplet is extremely thin and has a steep profile at the periphery of the JP as in the schematics of Fig.~\ref{fig:1}(a) and (c). 

%%this goes to Appendix 
%Since the angles $\theta_\mathrm{C}$ are material properties determined by the Young relations, the variation may originate from metastability by agitation when creating droplets. Note that $\theta_\mathrm{W}$ can be understood as the apparent angle because the JP is fully covered by the oil in this study. Therefore, no contact line pinning is expected. The JP has in-plane polarity in a random direction, unlike the interfacial trapped JPs, which have a propensity to have smaller in-plane polarity depending on the surface properties~\cite{Park2011,wang_enhanced_2015,Dietrich2017,dietrich2018active,dietrich_microscale_2020}. 

% \begin{figure}[tb]
% \includegraphics[width=0.5\textwidth]{figures/fig2_.pdf}
% \caption{\label{fig:2}Examples of regular and irregular motions of JP. (a, b) 
% Time-lapse images of a regular (a) and irregular (b) motion in a droplet under activation ($I=\SI{426}{\watt\per{\centi\metre}^2}$ for \SI{1.5}{\second} and $I=\SI{415}{\watt\per{\centi\metre}^2}$ for \SI{5}{\second}). The trajectories during the montages are shown. The droplet radii were $R=\SI{239.08}{\micro\metre}$ and $\SI{158.13}{\micro\metre}$ respectively. %(c, d) Trajectories of the regular and the irregular motions. The green circle means the illumination area. 
% See also Supplemental videos 1 and 2~\cite{supp_material}.
% (c, d) Positions $x(t),\,y(t)$,
% velocities $v_x(t),\,v_y(t)$, corresponding to (a, b). Red and blue lines are for $x$, $y$-component, and orange dashed lines are speeds $v(t)$.}
% % \end{figure}
\begin{figure}[tb]
\includegraphics[bb= 0 0 1750 1440,width=0.5\textwidth]{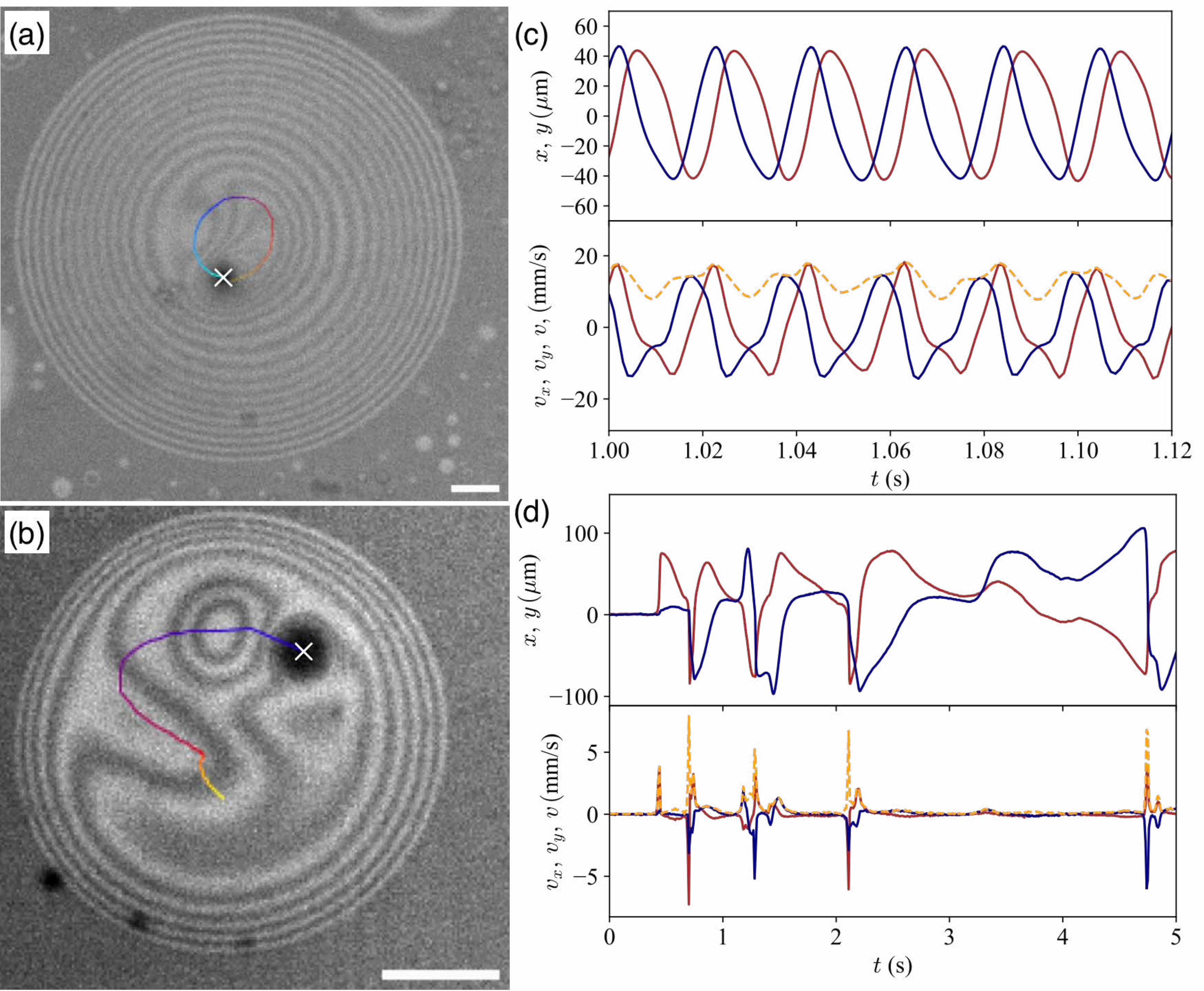}
\caption{\label{fig:2}Examples of regular and irregular motions of JPs. (a, b) 
Snapshots of a regular (a) and irregular (b) motion in a droplet under activation corresponding to $t=\SI{1.03}{\second}$ and $t=\SI{0.72}{\second}$ of (c) and (d) under illumination $I=\SI{426}{\watt\per{\centi\metre}^2}$ (a) and $I=\SI{415}{\watt\per{\centi\metre}^2}$ (b). The droplets have radii $R=\SI{239.08}{\micro\metre}$ and $\SI{158.13}{\micro\metre}$, respectively. The trajectories drawn are for a duration of \SI{20}{\milli\second} (a) and \SI{0.73}{\second} (b) and are directed from yellow to blue. The JP positions are marked by ``$\times$.'' %(c, d) Trajectories of the regular and the irregular motions. The green circle means the illumination area. 
See also Supplemental videos 1 and 2~\cite{supp_material}. The scale bars represent $\SI{100}{\micro\metre}$.
(c, d) Positions $x(t),\,y(t)$,
velocities $v_x(t),\,v_y(t)$, in the plane of the interface, corresponding to (a, b). The red and blue lines are for $x$, $y$-components, and the orange dashed lines indicate total speeds $v(t)$.}
\end{figure}

\smallskip
%about Fig.2
%\textit{Dynamic Regimes of JP motions.---}%
We now turn to how the thin confining droplets influence the motion of the self-propelled JP. We observe various planar JP fast motions with distinct temporal features using droplets with different sizes under different laser illumination intensities. To comprehend the variety of JP motions, we use the time series of the particle positions $x(t),\,y(t)$, and
velocities $v_x(t),\,v_y(t)$, in the plane of the interface.  Within the droplets, the JP can undergo either regular motions (circular, ellipsoidal, and occasionally back-and-forth motion) with clear periodicity or irregular motions with intermittency (see Fig.~\ref{fig:2} and Supplementary videos S1 and S2~\cite{supp_material}). In the regular regime, the JP can maintain a periodic motion with a typical velocity of \SI{1}{\centi\metre\per\second} for more than \SI{17}{\second} (within the limit of our observation time). In contrast, in the irregular regime, the motion of the JP is intermittent, and the trajectory becomes rather chaotic. 

%In the regular regime, JP can maintain a circular-like motion with periodic planar velocities $v_x$ and $v_y$ hovering between $-20$ \SI{}{\milli\metre\per\second} and $20$ \SI{}{\milli\metre\per\second}. 

% In the regular regime, JP went to a periodic motion (limit cycle) with a circular-like trajectory at almost periodic velocity after some transient time. The periodic state lasted more than \SI{17}{\second} within the limit of the observation time. In contrast, in the irregular regime, the JP moved irregularly from the beginning of the illumination and never showed apparent periodicity in velocity, though the JP was kept confined in the droplet. 

\begin{figure*}[tb]
\includegraphics[bb= 0 0 2611 697,width=\textwidth]{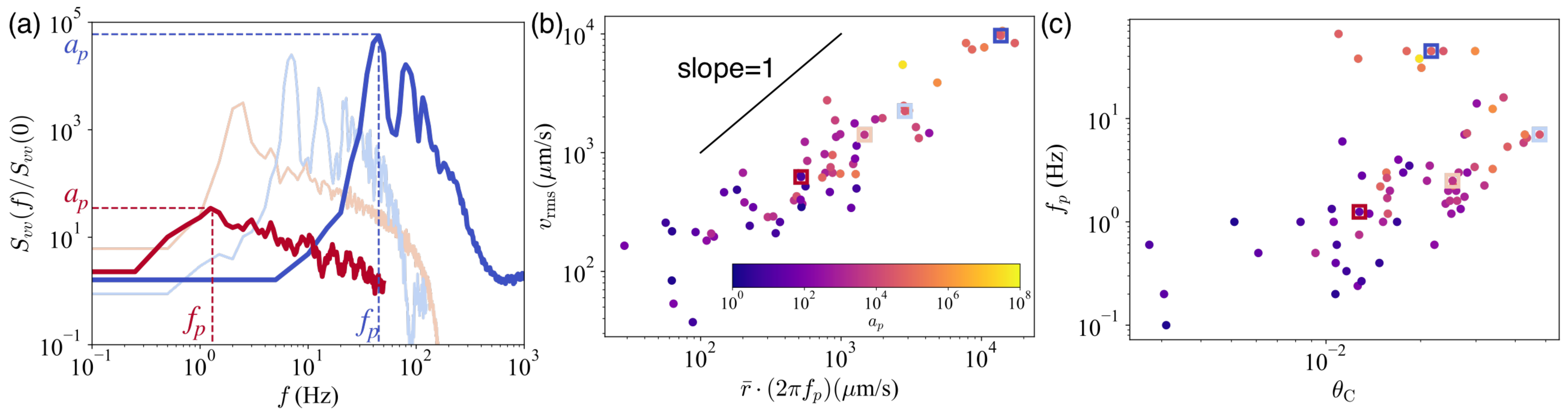}
\caption{\label{fig:3} Analysis of various trajectories in steady states. (a) Typical examples of PSDs. Regular (blue): $I=\SI{426}{\watt\per{\centi\metre}^2}$ corresponding to Fig.~\ref{fig:2} (a,c); irregular (red): $I=\SI{415}{\watt\per{\centi\metre}^2}$ corresponding to Fig.~\ref{fig:2} (b,d). The other two curves represent intermediate cases in terms of the peak heights ($I=\SI{565}{\watt\per{\centi\metre}^2},\,\SI{511}{\watt\per{\centi\metre}^2}$).  The characteristic frequencies $f_p$ and corresponding amplitudes $a_p$ of the main peaks of the spectra are indicated by dashed lines. (b) Relationship between the RMS velocity $v_\mathrm{rms}$ and $\bar{r}f_p$. (c) Correlation between the frequency $f_p$ and the contact angle $\theta_\mathrm{C}$. In (b, c), the corresponding cases of (a) are marked by the square of the same colors. %(c) Phase diagram as a function of $(P,\tan^2\theta_\mathrm{C}$) colored by $v_\mathrm{rms} (\propto f_\mathrm{p}$; regularity). The symbol ``$\times$'' means no significant motion was observed. The inset shows that the time-averaged radii $q$ is constant, showing no correlation with the droplet size $R$. The solid black line with slope 1 is a guide to the eye. 
}
\end{figure*}
\smallskip
%about Fig.3
%\textit{Characterization of the regularity and its origin.---}%
%\textit{Origin of the regularity.---}
% steady state 
% exclude the obvious change
% Here, we focus on the temporal features of the regularity of JP dynamics in the steady states. We applied the power spectral density (PSD) of velocities: $S_{vv}(f):=|\hat{v_x}(f)|^2+|\hat{v_y}(f)|^2$, where $\hat{v_j}(f):=\frac{1}{N}\sum_{k=0}^{N-1} v_j(t_k) e^{-2\pi i f t_k}\, (j=x,y)$, $t_k$ is the time at the $k$th video frame, $f$ is the frequency, and $N$ is the total frame number. Note that we here only focus on the droplets without apparent radii changes in the course of JP motion (see SI). We characterize the regularity of dynamic regimes by using the peak frequency $f_\mathrm{p}$ in the spectrum (Fig.~\ref{fig:3}(a)). 
To characterize the observed particle trajectories, we examine the power spectral density of the velocity (PSD): $S_{vv}(f):=|\hat{v_x}(f)|^2+|\hat{v_y}(f)|^2$, where $\hat{v_j}(f):=\frac{1}{N}\sum_{k=0}^{N-1} v_j(t_k) e^{-2\pi i f t_k}\, (j=x,y)$. Here $f$ is the frequency and $t_k$ is the time with $N$ being the total number of measurements. Note that we here focus on the steady-state behaviors of JPs within droplets with a roughly constant radius during the entire trajectory.  
%Due to the confinement in a droplet during motions, we can extract typical frequency even for highly irregular motions. 
The power spectra of the velocity time series of different trajectories show a prominent peak at a fundamental frequency $f_p$ with a relative amplitude $a_p:=S_{vv}(f_p)/S_{vv}(0)$ as shown in Fig.~\ref{fig:3}(a), satisfying $a_p\sim f_p^2$ (see details in \cite{supp_material}). However, two types of spectra can be distinguished: regular trajectories generally give rise to spectra with well-defined sharp peaks with large amplitudes, while irregular motions give rise to broad spectra with smaller amplitudes. We find a relationship $v_\mathrm{rms} \approx \bar{r} \cdot (2\pi f_p)$, which is valid for both regular and irregular motions across all our observations. Here, $v_\mathrm{rms}:=\overline{\sqrt{|{\bf v}(t)|^2}}$ represents the time-averaged speed during steady states, while $\bar{r}$ represents the mean radius of the trajectory.
Each JP--droplet combination has different values of $I,\, R,\,\theta_\mathrm{C}$, the contact angle, and $\theta_\mathrm{W}$, the apparent contact angle near the particle, raising the question of which parameters are essential in determining whether the JP motion is regular or irregular.
Here, we further find a correlation between the frequency $f_p$ and the contact angle $\theta_C$ as shown in Fig.~\ref{fig:3}(c).

\smallskip
%about Fig.4
%\textit{Activation mechanism and oil thickness-coupled Janus particle motion.---}%
Similarly to light-driven JPs at a planar single interface \cite{dietrich_microscale_2020,wurger_thermally_2014}, we believe that our JPs are driven by a thermal Marangoni flow at speed $v_p \sim -\mathrm{d}\gamma/\mathrm{d}T\cdot \frac{\Delta T}{\eta_o}$, where $\gamma$ is the interfacial tension, $T$ is the temperature, $\Delta T$ is the temperature rise due to light absorption by the metal cap and $\eta_o$ the viscosity of the liquid. This results in velocities of $\mathcal{O}(\SI{1}{\centi\metre\per\second})$ for the air--oil interface and $\Delta T\sim\SI{1}{\kelvin}$ with the particle moving with its metal cap facing forward as the Marangoni flows go from the hot to the cold region. 

% More precisely, in thin oil film, the speed is estimated as $v_p=-\frac{bu_W}{a^2}\cdot\frac{\mathrm{d}\gamma}{\mathrm{d}T}\cdot \frac{\Delta T}{3\eta_o}$, where $b$.

% The temperature dependences of the interfacial tensions are followings: $\frac{d\gamma_\mathrm{O}}{dT}=-0.144 \SI{}{\milli\newton\per\metre\per\kelvin}$ (air--oil)~\cite{Queimada2001}, $\frac{d\gamma_\mathrm{OW}}{dT}=-0.08 \SI{}{\milli\newton\per\metre\per\kelvin}$ (oil--water)~\cite{Zeppieri2001}.
%  Using the description in Dietrich {\it et. al.}~\cite{Dietrich2020}, the temperature rise might be up to several K.
To support this, we conducted similar experiments by replacing a JP with a gold core-shell particle~\cite{supp_material}. Heating by laser illumination changed the fringe pattern symmetrically around the particle, but resulted in no significant motion. We thus believe, the polarity of the JP is essential, unlike the case of reference~\cite{Koleski2020}. In addition, the JP generated flow only locally, as suggested by additional experiments using tracer particles (see Sec.~XIV~\cite{supp_material} and Supplemental Video 3). %This propulsion mechanism differs greatly from the instability of global flow generation by a heated symmetric particle reported in.

Previous studies on thermal Marangoni surfers at an interface showed that the particles propelled at constant speed $v_p \propto \Delta T/\eta_o \propto I$ both at planar ~\cite{dietrich_microscale_2020,wurger_thermally_2014} and curved interfaces~\cite{ben_zion_cooperation_2022, ganesh_dynamics_2023}. In contrast, the speed of our JPs changes in time (see Fig.~\ref{fig:2}) through the coupling of the JP motion and the interfacial profile, as seen through the changes of the fringe pattern in Fig.~\ref{fig:2}(a, b) and Supplementary videos S1 and S2~\cite{supp_material}. Notably, the coupling occurs due to the thin nature of the oil droplets, not due to the intrinsic curvature of the droplet, as demonstrated in our thin film experiments where $v_p\propto I$ is valid only for a thick film $h\approx 12a$, not in a thin film $h\approx 2a$ (see Sec.~VIII of \cite{supp_material}).

%start theoretical part from here. try to explain physical origin
The coupling ensues as particle motion induces interfacial deformation, and the resulting deformation, in turn, influences the particle dynamics. The physical mechanism of the deformation is due mainly to asymmetric thermal Marangoni flow driving the JP, which hydrodynamically affects the lens deformation due to the finite volume of the oil phase, together with the effect of the moving solid boundary of the JP. This deformation is unique to the non-equilibrium state and can give rise to a local capillary force. In addition, this deformation can yield a net torque on the particle because of the rotational asymmetry of the deformation around the JP, unlike the trivial deformation due to the existence of an off-centered passive particle thicker than the film. 

%Here, in turn, we are going to verify the opposite effect: what deformation is really produced and how it affects the JP motion.
%The coupling of the motion of the particle to the thickness profile changes is very complex when heating is present due to fluid flow including thermal Marangoni flow and the presence of forces exerted on the JP and due to the time-dependent inhomogeneous thickness profile. This force can be understood as a type of capillary force acting on the particle~\cite{sur_capillary_2001,yadav_capillary-induced_2019} when it is embedded in a thin film with thickness gradients present.

The local capillary force has been proposed in reference~\cite{yadav_capillary-induced_2019} and is rewritten here in its two-dimensional version:
\begin{equation}\label{eq:yadav}
    {\bf F}({\bf r})=-\frac{\pi\gamma (h({\bf r})-h_e)}{1+(\frac{\partial h}{\partial {\bf r}})^2}\nabla h,
\end{equation}
where ${\bf r}$ is the position vector and $h_e$ is the equilibrium wetting thickness for the particle. 
%As we mentioned above, the centripetal force observed can be related to this capillary force.  

\begin{figure*}[tb]
\includegraphics[width=\textwidth]{figures/fig4.pdf}% Here is how to import EPS art
\caption{\label{fig:4} Capillarity-induced circular motion. (a) Reconstructed thickness profile of a snapshot of the drop with the particle undergoing circular motion ($f_p\approx45$Hz, $I=\SI{565}{\watt\per{\centi\metre}^2}$). The black dot and the arrow indicate the particle position and the propelling direction. The center of the droplet is marked as $\times$. Inset: Schematic for the capillary force and torque exerted on the JP in a thin droplet as well as the cap orientation $\hat{\bf p}$. (b) Examples of the time evolution of position $\tilde{x}(\tilde{t}),\,\tilde{y}(\tilde{t})$ and torque $A\tilde{\bf r}\cdot \hat{\bf p}$ (red, blue and green in the top figure). The corresponding $\tilde{v_x}(\tilde{t}),\,\tilde{v_y}(\tilde{t})$, and speed $\tilde{v}(\tilde{t})$ (orange) are in the bottom figure. For this case, $A=10$, ${\bf r}|_{t=0}=(0.1,0)$, and $\hat{\bf p}|_{t=0}=(1/\sqrt{2},1/\sqrt{2})$ are used. (c) The limit cycles for different parameters. (d) Steady-state values of the speed $v^\mathrm{SS}$, frequency $f^\mathrm{model}$ and the radial length $\bar{r}^\mathrm{model}$ for each parameter $A$ (normalized). }
\end{figure*}

We utilize Eq.~\eqref{eq:yadav} and estimate this capillary force using the reconstructed droplet thickness profile during the steady circular motion. A snapshot of such a profile is shown in Fig.~\ref{fig:4}(a). The inhomogeneous profile around the JP results in a nonzero net force ${\bf F}^{cap}$ and torque $T^{cap}$ exerted on the JP as defined in the inset of Fig.~\ref{fig:4}(a). 
To estimate the force and torque, we assume $h_e:=2u_\mathrm{w}$ at $r=r_\mathrm{w}$ (see Fig.~\ref{fig:1}(c)): the estimated value of wetting thickness without laser illumination. We take a force application line cycle $C_\zeta$ with radius $\zeta$ centered at the JP. The net force and torque are calculated as ${\bf F}^{cap}_\zeta=\oint_{C_\zeta} {\bf F}(s)\,\mathrm{d}s$ and $T^{cap}_\zeta=\oint_{C_\zeta} \zeta{\bf F}(s)\cdot {\bf t}\,\mathrm{d}s$, where $\mathrm{d}s$ and ${\bf t}$ are the infinitesimal length elements along the curve $C_\zeta$ and the tangential unit vector along $C_\zeta$. Since the force field decays with $\zeta$~\cite{supp_material}, the average within $r_w\leq \zeta \leq \zeta^\mathrm{max}$ was taken, where $\zeta^\mathrm{max}$ is a distance above which the forces and torques are negligible. Then, we can obtain the net force ${\bf F}^{cap}$ and net torque $T^{cap}$. For more details, see Sec.~IX and Table S1~\cite{supp_material}. We obtained capillary centripetal forces of the order of $\SI{10}{\nano\newton}$, which are comparable to the viscous forces $F_v=\xi v_p$ ($\xi=6\pi a \eta_o$). 

Also, there is the capillary positive azimuthal force that pushes the JP forward at the same order of the centripetal force. We often see a thinning at the tail of the JP, as seen also in Fig.~\ref{fig:2}(a,b), so the rear profile becomes steeper, which pushes the JP forward. 

Previously, Dauchot {\it et al.} showed that an active particle subject to a generic self-aligning torque and immersed in a harmonic trap ~\cite{Dauchot2019} can exhibit different types of trajectories such as orbiting trajectories or so-called climbing trajectories as well as a coexistence between them depending on inertia and the self-alignment intensity. Although the coexistence region occurs only in the presence of sufficient inertia which is absent in our experiments, some of our observations are in line with this model in particular for the orbiting motion in the capillary confinement. In the scenario of Dauchot {\it et al.}~\cite{Dauchot2019}, the particle feels a torque due to the generic self-alignment of the particle. In contrast, the origin of the torque in our case is capillarity and can be estimated experimentally from the dynamic deformation of the drop in the vicinity of the particle. The capillary torque $T^{cap}$ is calculated from the observed profile via \eqref{eq:yadav}, and turns out to be comparable to the viscous torque (see Table S1 in \cite{supp_material}). For a particle with counterclockwise circular motion, a positive torque represents an inward rotation of the JP polarity. Note that the nonzero torque $T^{cap}$ means that the capillary force field is not conservative. This is partially because the force (Eq.~\eqref{eq:yadav}) is nonlinear. In addition, the equilibrium thickness $h_e$ in Eq.~\eqref{eq:yadav} is not necessarily constant and may vary spatiotemporally due to the fluid flow and the asymmetric heating, for example.
Therefore, the circular motion can be understood by the combination of the capillary centripetal force and the capillary positive torque, as well as a self-propulsion.

Here, to better understand the mechanism of the steady circular motion more quantitatively, we consider the following model with an overdamped equation of motion for the particle subjected to rotation by capillarity:
\begin{eqnarray}
    0=-\xi(\dot{\bf r}-v_p \hat{{\bf p}})-k{\bf r}\label{eq:eom}\\
    \frac{\mathrm{d}\hat{{\bf p}}}{\mathrm{d}t}=\bf{\Omega}\times\hat{{\bf p}}\label{eq:polarityrotation}\\
    0= {\bf T}^{cap}-\xi_r {\bf \Omega}\label{eq:eomrotation},
\end{eqnarray}
where ${\bf r}$ is now the position of the particle in the plane, ${\bf \Omega}$ and $\hat{\bf p}$ are the angular velocity of the JP self-rotation and the JP polarity (cap orientation). Equation~\eqref{eq:eom} takes into account that the particle is powered by the self-propulsion force $F_\mathrm{sp}:=\xi v_p$, whose value can be shifted by nonzero $F_\theta^{cap}$. The particle also feels the viscous damping and the capillary force in the direction ${\bf r}$ or centripetal force which is approximated, for simplicity and ease, as $-k\bf r$ with $k$ given by 
\begin{equation}\label{eq:spring}
    k=4\pi \gamma \tan^2 \theta_\mathrm{C},
\end{equation} 
derived via the excess interfacial energy by the lowest order deformation of the droplet profile (see the derivation~\cite{supp_material}). This expression matches with Eq.~\eqref{eq:yadav}
in the limit where the profile is considered locally linear. The centripetal force $|k\bar{r}|\sim \mathcal{O}(\SI{10}{\nano\newton})$ using the measured $\theta_\mathrm{C}$ is consistent with the estimation from the profile ($F^{cap}_r$). Experimentally, the capillarity-induced centripetal force can be determined via the relaxation dynamics soon after switching off the laser illumination (see Sec.~VII ~\cite{supp_material}). Neglecting inertia and the azimuthal motion, this force balances the viscous friction, allowing to obtain direct estimates of the value of the spring constant $k$. 

Equations \eqref{eq:polarityrotation} and \eqref{eq:eomrotation} take into account the capillary torque ${\bf T}^{cap}$ which can rotate the particle and thus its polarity. This torque is balanced by viscous damping with a rotational damping coefficient $\xi_r$.  
Here we may assume that the capillary torque satisfies ${\bf T}^{cap}=\Gamma ({\bf r}\times \hat{{\bf p}})$ with the coefficient $\Gamma>0$ for the observed positive torque. It can be natural to assume the magnitude of the capillary torque to be proportional to that of the capillary force and zero torque for a strictly radial $\hat{\bf p}$.
The Eqs.~(\ref{eq:eom}--\ref{eq:polarityrotation}) are recast as the following nonlinear equations with the non-dimensionalized time  $\tilde{t}:=\tau t$, and position $\tilde{\bf r}:={\bf r}/v_p \tau$  with a single control parameter $A:= \frac{v_p \tau^2 \Gamma}{\xi_r}$, where $\tau =\xi/k$:
% The Eqs.~(\ref{eq:eom}--\ref{eq:polarityrotation}) are summarized as follows with the non-dimensionalization, we obtain
\begin{eqnarray}
    \frac{\mathrm{d}\hat{\bf p}}{\mathrm{d}\tilde{t}}=A \left\{ (\tilde{\bf r}\cdot \hat{{\bf p}})\hat{\bf p}-\tilde{\bf r} \right\}\label{eq:p_nodim}\\
    \frac{\mathrm{d}\tilde{\bf r}}{\mathrm{d}\tilde{t}}=\hat{\bf p}-\tilde{\bf r}\label{eq:r_nodim}.
\end{eqnarray}

% \begin{eqnarray}
%     \frac{\mathrm{d}\hat{\bf p}}{\mathrm{d}\tilde{t}}=A \left\{ (\tilde{\bf r}\cdot \hat{{\bf p}})\hat{\bf p}-\tilde{\bf r} \right\}\label{eq:p_nodim}\\
%     \frac{\mathrm{d}\tilde{\bf r}}{\mathrm{d}\tilde{t}}=\hat{\bf p}-\tilde{\bf r}\label{eq:r_nodim}.
% \end{eqnarray}
The non-dimensionalized equations are solved using a Python code. The numerical results are shown in Fig.~\ref{fig:4}(c,d). Unless the initial polarity is strictly radial, the dynamics results in steady states with either periodic circular motions at a frequency $f^\mathrm{model}$ with constant torque and speed for $A\geq 1$, or in static states without any motion for $A < 1$. In fact, the static states were observed experimentally for a small laser illumination (see Fig.~S2~\cite{supp_material}). The steady circular trajectories, namely, the limit cycles, are shown in Fig.~\ref{fig:4}(c). Large torques, fast propulsion velocities $v_p$, or large values of $\tau$ all result in large values of $A$, leading to rotation with a smaller radius $\bar{r}$. The steady-state speed $v^\mathrm{SS}:=\bar{r}^\mathrm{model}\cdot (2\pi f^\mathrm{model})$ for periodic motions (set to zero for static states), is shown in Fig.~\ref{fig:4}(d). The speed, comparable to $v_p$, is consistent with our experimental observations. In addition, the numerical solution leads to the relations $f^\mathrm{model}\propto \sqrt{\frac{v_p \Gamma}{\xi_r}}$ and $\bar{r}^\mathrm{model}\propto \sqrt{\frac{v_p\xi_r}{\Gamma}}$. Therefore, a larger capillary torque $\Gamma$ results in motions at a higher frequency $f^\mathrm{model}$. Considering the experimental observation that a larger $\theta_\mathrm{C}$ leads to higher-frequency motion ($f_p$) as shown in Fig.~\ref{fig:3}(c), $\Gamma$ must increase with $\theta_\mathrm{C}$. Since $f^\mathrm{model}\sim \mathcal{O}(\qtyrange{10}{100}{\hertz})$ for $\tau\sim 10^{-2}\SI{}{\second}$ (for the regular cases), the frequency range turns out to be consistent with our measurements. 

%\textcolor{blue}{though the radius can change due to the droplet confinement and limited laser illumination area????Though the magnitude of the capillary torque $\Gamma$ is not quantitatively connected with that of the capillary force $k$, the typical frequency ($f_p$ or $f^\mathrm{model}$) increases with the larger capillarity both experimentally and theoretically.}.  \textcolor{red}{how to connect $f_p\propto k$ and $f^\mathrm{model}\propto A^{1/2}$ more conclusively? } 

This model is consistent with the model proposed by Dauchot {\it et al.}~\cite{Dauchot2019} describing circular motions. 
However, as we have already pointed out, the speed is not constant but periodic in our circular motions. The thickness profiles at different times during circular motion shown in Fig.~S7~\cite{supp_material} (a, b) are similar but not the same, the approximation of the constant capillary force and torque is not perfect. The difference in the azimuthal capillary forces results in the speed modulation in time. We believe the differences in thickness profiles with time are present because the time scale for oil film relaxation is comparable to that of the particle dynamics.

In our experimental observations, the irregular state occurs at even smaller values of $k$ than those of circular motions. These irregular motions are far beyond the scope of our model. The approximation that the capillary effects can be modeled as a constant centripetal force and a constant torque most probably breaks down in this case because the thickness profile changes chaotically with the particle motion. The dynamic film deformation must give rise to a time-dependent potential $U({\bf r},t)=kr^2/2 +U'({\bf r},t)$ where $U'(t)$ changes in time (see discussion in Sec.~VII~\cite{supp_material} and Supplemental Video 4~\cite{supp_material}). Other limitations of the model arise. The non-conservative capillary torque is created by asymmetric Marangoni flow, but fully determining this deformation requires solving the coupled particle dynamics, hydrodynamics, and temperature field, which is beyond the scope of this work. Further, the existence of the solid particle poses challenges to the thin-film hydrodynamic approaches commonly used in related studies~\cite{Trinschek2020,Clavaud2021}.

We have observed additional intriguing features such as back-and-forth motions. These may be understood as transient states to a circular motion or a result of out-of-plane rotation of the polarity (see discussion in Sec.~VIII~\cite{supp_material}). Moreover, it is worth noting that, in a transient case with an apparent planar dilation (increase in $R$), periodic circular motion turned to irregular motion suddenly (see Supplemental Video 5~\cite{supp_material}). This suggests that the decrease in $\theta_\mathrm{C}$ due to the dilation and the oil volume conservation may have driven the system into an irregular state as the droplet became much thinner highlighting the role of temporally changing capillary forces and torques in setting the type of particle dynamics.

%Moreover, it is worth noting a transient case where a clear planar dilation (increase in $R$) was observed: initially, the JP exhibited periodic circular motion, but as the dilation progressed, its trajectory became irregular at a certain point (see Supplemental Video 3~\cite{supp_material}). This suggests that the decrease in contact angle $\theta_\mathrm{C}$ due to the oil volume conservation may have driven the system into the irregular state.

%In fact, we have observed a variety of cases, including large deformations (see Supplemental video 4~\cite{supp_material}) and even an extreme case for which the JP ends up escaping from the oil phase in an extremely thin droplet, possibly by film breakage or dewetting~\cite{fiegel_wetting_2005} (see Supplemental video 5~\cite{supp_material}). 

\smallskip
%\textit{Concluding remarks.---} 
In conclusion, we observed novel dynamics of active Janus particles confined in thin interfacial droplets. The coupling between droplet thickness and particle motion induces circular motion and other non-trivial trajectories due to the strong confining geometry. We proposed a simple model where capillarity plays a major role in setting the properties of particle trajectories. Insights from studying active particles in thin films could provide a crucial understanding for developing novel active counterparts of particle-based soft matter systems, such as suspension films, capillary suspensions, and Pickering emulsions. The interplay between interfacial properties and activity is expected to yield unique characteristics in activity-controlled functional materials, where manipulating activity offers new avenues for creating smart and responsive systems.

\medskip
\begin{acknowledgments}
We acknowledge the support of the French Agence Nationale de la Recherche (ANR), under grant N° ANR-22-CE06-0007-02.
We thank T. Bickel, Y. Tagawa, K. A. Takeuchi for helpful and interesting discussions. We are particularly indebted to A. W\"{u}rger for numerous discussions and for his calculation of the drop profile as well as the centripetal force. 
\end{acknowledgments}

\end{document}

%\preprint{APS/123-QED}

\title{Supplementary information for ``Active particle in a very thin interfacial droplet''}% Force line breaks with \\
\author{Airi N. Kato$^{1}$}%
%\email{airi.nakamoto@u-bordeaux.fr}
\author{Kaili Xie$^{1,2}$}
\author{Benjamin Gorin$^{1}$}
\author{Jean-Michel Rampnoux$^{1}$}
%\author{Alois W\"{u}rger$^{1}$}
\author{Hamid Kellay$^{1}$}%
 \email{hamid.kellay@u-bordeaux.fr}
\affiliation{
$^{1}$ Laboratoire Ondes et Mati\`ere d’Aquitaine, Universit\'e de Bordeaux, Talence 33405, France
}%
\affiliation{%
 $^{2}$Van der Waals-Zeeman Institute, Institute of Physics, University of Amsterdam, 1098XH Amsterdam, The Netherlands
}%
%\author{Airi N. Kato}
%\email{amo.rationem@gmail.com}
%\affiliation{%
% Wenzhou Institute, University of Chinese Academy of Sciences,
%No.1 Jinlian road, Longwan district, Wenzhou, Zhejiang province, China 325001}%
%\address{}

\maketitle

\section{Preparation of Janus particles}
\begin{figure}[h]
\includegraphics[width=6cm]{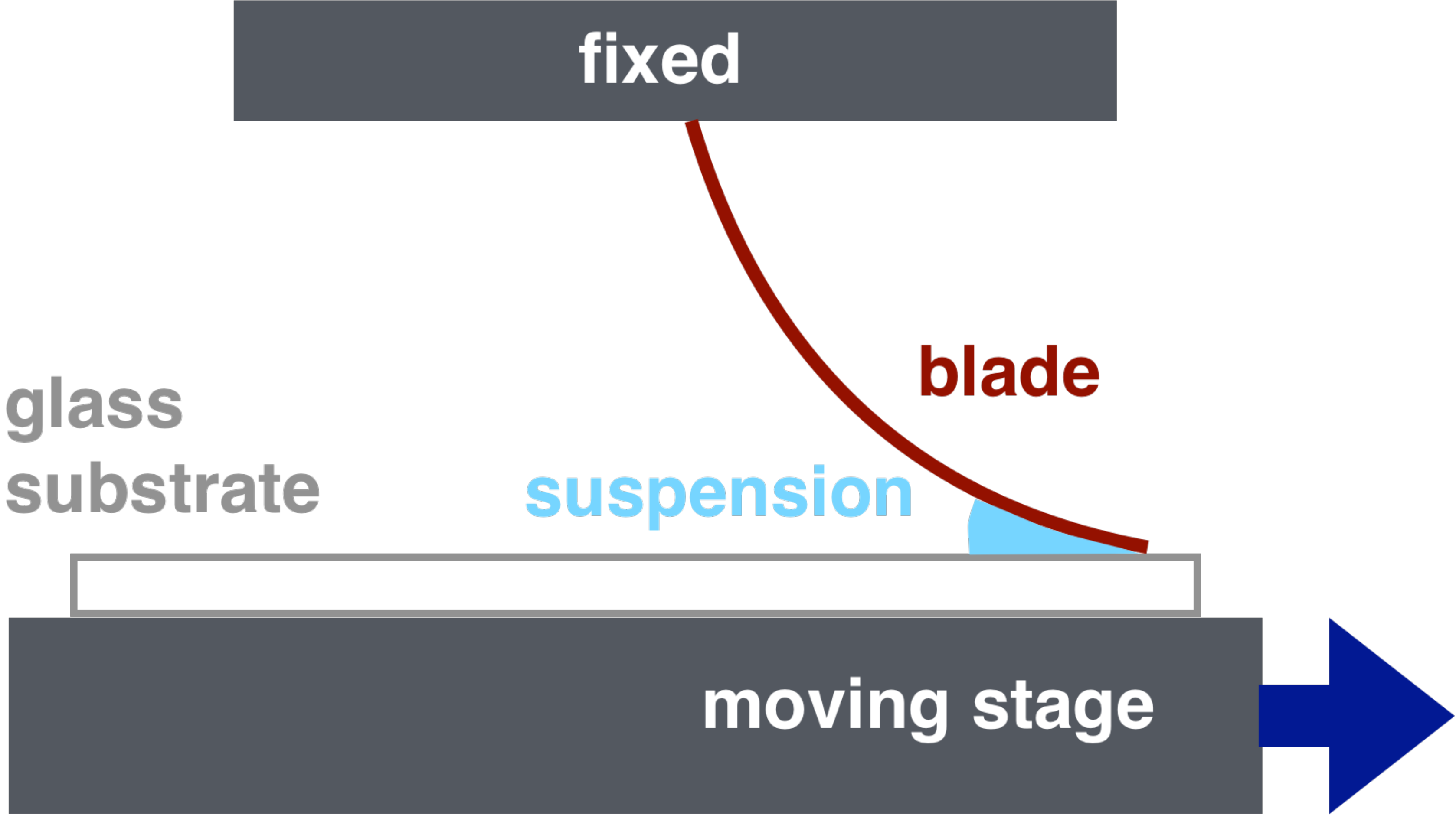}
\caption{\label{fig:bladecoating} Schematics of blade coating of bare particles to make a monolayer. }
\end{figure}

We prepared Janus particles (JPs) by metal coating polystyrene particles which were previously deposited as a monolayer on clean glass substrates. 
The polystyrene beads (TS10, Dinoseeds) were washed by Milli-Q twice and re-suspended in Milli-Q at a volume of approximately 50\%. 
The glass substrates are washed with soap water, rinsed by Milli-Q, and treated with a plasma cleaner for $15$ minutes after drying to make them more hydrophilic.
The particle suspension was deposited using a flexible blade to make a particle monolayer on the clean glass as illustrated in Fig.~\ref{fig:bladecoating}, after the sonication of the suspension.

After drying, \SI{10}{\nano\metre} chromium layer followed by a \SI{100}{\nano\metre} gold layer were subsequently deposited on the particles using an electron beam physical vapor deposition technique (EB-PVD, Meca2000).
The coated particles were removed from the substrates by sonication in a water-ethanol mixture, washed with Milli-Q water several times, and dried at 35 degrees for a few days.
%(Orca Flash 4.0 from Hamamatsu) 
% these details may go to SI: We observed this composite system by the reflection of green light (green-filtered white LED) via microscopy (Axio Observer, Zeiss) with x5 or x10 objective lens (N-Achroplan 5x/0.13, N-Achroplan 10x/0.25 Ph1 M27) and recorded using a highspeed camera (v641, Phantom).

\section{Droplet thickness profiles without laser activation}
Lens-shaped oil droplets can form at an air--water interface when $|\gamma_\mathrm{\it o}-\gamma_\mathrm{\it ow}|<\gamma_\mathrm{\it w}<\gamma_\mathrm{\it o}+\gamma_\mathrm{\it ow}$, satisfying the Young relation for three fluids:
\begin{eqnarray}
\gamma_\mathrm{\it w}&=&\gamma_\mathrm{\it o}\cos{\theta_1}+\gamma_\mathrm{\it ow}\cos\theta_2\nonumber\\
\gamma_\mathrm{\it o}\sin\theta_1&=&\gamma_\mathrm{\it ow}\sin \theta_2,\label{eq:Young}
\end{eqnarray}
where $\gamma_{\it o},\,\gamma_{\it ow},\,\gamma_{\it w}$ are the interfacial tensions of air--oil, oil--water, and air--water interfaces. In our cases, the values shown in Table \ref{table:materialint} hold lens-shaped oil droplet conditions, exhibiting small contact angles $\theta_{1}\approx 0.022$ and $\theta_{2}\approx 0.013$ by solving Eq.~\eqref{eq:Young}.

\begin{figure}[h]
\includegraphics[width=17cm]{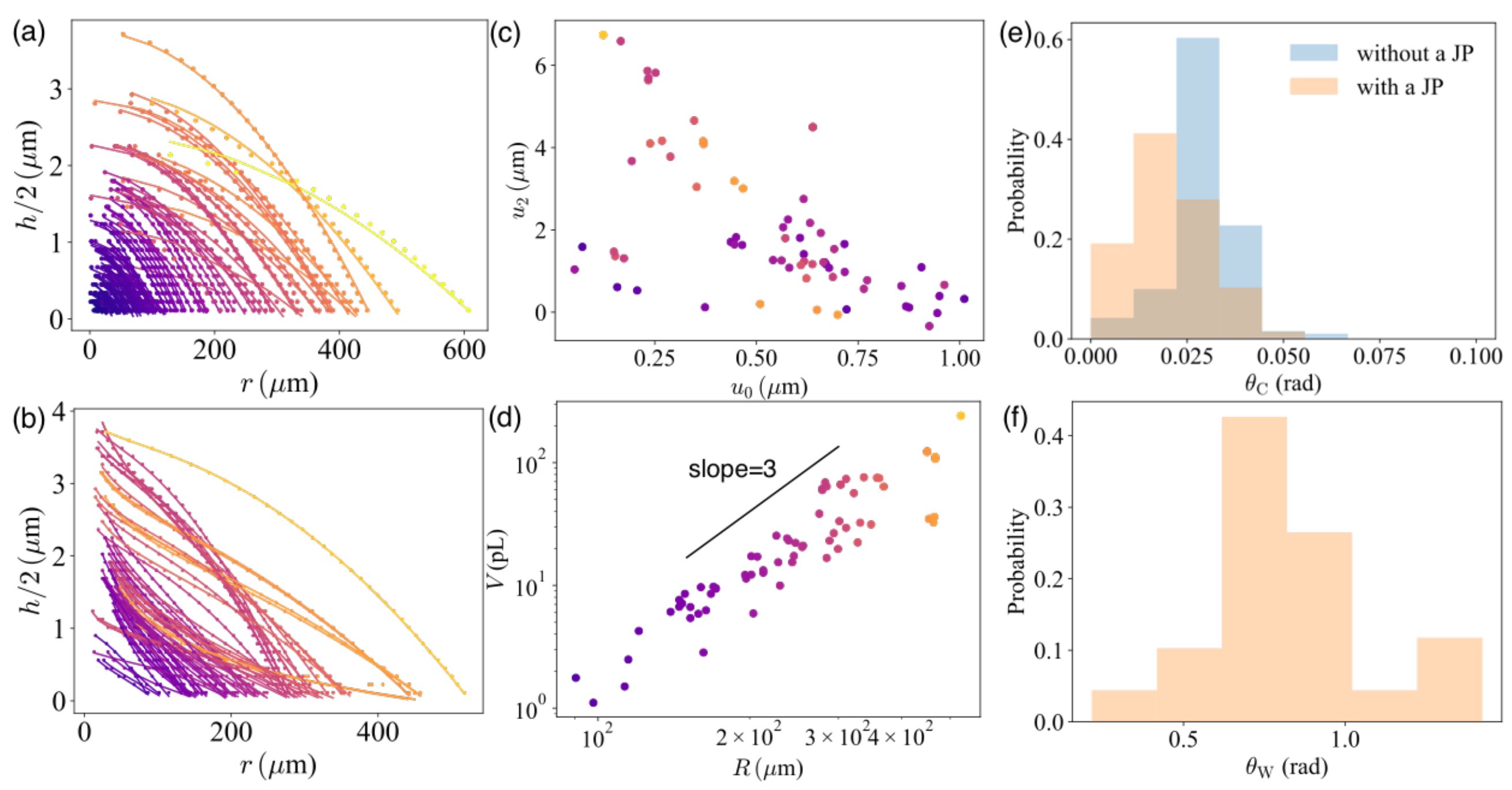}
\caption{\label{fig:S1} (a) Interfacial profiles of particle-free droplets. Dots are from the observation of the bright rings, with fitted solid lines using the second term of Eq.~\eqref{eq:u}. (b) Interfacial profiles of droplets with a passive JP. Dots are from the observation of the bright rings, with fitted solid lines using Eq.~\eqref{eq:u}. The colors from purple to yellow of (a,b) are set by $R$. (c) The two fitting parameters for the fitting to (b). (d) The estimated volume $V$ by substituting the fitting parameter $u_0,\,u_2$ to Eq.~\eqref{eq:V} for fixed droplet radius $R$. (e) The probability distribution of contact angles $\theta_\mathrm{C}$ for the free drops and the droplet with the JP. (f) The probability distribution of the wetting angle $\theta_\mathrm{W}$.}
\end{figure}

Next, we discuss the profiles of oil droplets and those enclosing the immobile JP at the interface. %The oil film ranges to a maximum radius of $R$.  
For the sake of simplicity, we assume the oil droplet is symmetric with $\theta_1 = \theta_2$ (see Fig.1 in the main text). The interfacial profile $u(r)$ is defined as $u(r)=h(r)/2$, where $h(r)$ is the oil thickness. The profile $u({\bf r})$ satisfies the Young--Laplace equation
\begin{equation}
    \gamma \nabla^2 u+ p_0 =0,
\end{equation}
with the interface tension $\gamma$ and the internal pressure $p_0$ in the oil phase. Putting $u_2 = p_0R^2/\gamma$, one readily
finds
\begin{equation}\label{eq:u}
    u(r)=-u_0\ln\frac{r}{R}+u_2 \frac{R^2-r^2}{2R^2}
\end{equation}
The three parameters $u_0,\, u_2,\, R$ are determined by the droplet volume $V$, and the
contact angles $\theta_\mathrm{c}(=\theta_{1}+\theta_2)$ and $\theta_\mathrm{w}$ at the air--water--oil phase boundary and where the fluid interfaces meet the particle surface. 
For the particle-free droplet $u_0=0$, the second term of Eq.~\eqref{eq:u} gives the quadratic interfacial profile. As explained in the main text, we reconstruct the thickness profile from the fringes for particle-free droplets, and droplets with a passive JP are shown in Fig.~\ref{fig:S1}(a, b) with the fitting to Eq.~\eqref{eq:u}. Here we fixed $R$ to the measured droplet radius and conducted the two-parameter fitting. The fitted parameters are shown in Fig.~\ref{fig:S1}(c).

The droplet volume is readily integrated Eq.~\eqref{eq:u}:
\begin{equation}\label{eq:V}
    V=\frac{u_0 R^2}{4}+\frac{u_2 R^2}{8},
\end{equation}
where we neglect terms of the order $\mathcal{O}(a^2/R^2)$, and shown in Fig.~\ref{fig:S1}(d) as a function of $R$. The volume $V$ is roughly proportional to $R^3$, and the droplet volumes can range \SI{1}{\pico\litre}--\SI{100}{\pico\litre}. As the volume of a JP is \SI{0.65}{\pico\litre}, the volume fraction of the interfacial droplets can widely range from 0.65\% to 65 \%.

By the full function shape with the fitting parameters $u_0,\, u_2$, the contact angles $\theta_\mathrm{C}$ and wetting angles $\theta_\mathrm{W}$ can be calculated. The probability distributions of the angles for all droplets are shown in Fig.~\ref{fig:S1}(e) and (f). The three-phase contact angles $\theta_\mathrm{C}$ slightly differ with larger variations $\theta_\mathrm{C}\simeq 0.58^\circ\pm 0.30^\circ $ with a JP case, compared to the cases of free droplet $\theta_\mathrm{C}\simeq 0.84^\circ\pm 0.27^\circ$. The wetting angle $\theta_\mathrm{W}$ significantly smaller than 90 degrees suggests the sharp interfacial thickness changes around the particle. 

\section{Laser beam for activation}
The details of our setup for creating a top-hat laser beam was described in our previous paper~\cite{xie_activity_2022}. In short, the Gaussian green laser beam (MSL532 \SI{300}{\milli\watt}, \SI{532}{\nano\metre} wavelength) is transformed into a homogeneous top-hat profile with \SI{108}{\micro\metre} radius by a beam shaper (TOPAG FBSR-20-532).

\section{Derivation of the spring constant of capillary centripetal force}%\textcolor{red}{Alois's derivation.}
In the lubrication approximation, the shear stress $\sigma=\eta v/u$ (where $\eta$ and $u$ are the viscosity of the oil film and the interface height) needs to be finite, imposing zero velocity at $r=R$. Thus, one expects that the boundary of the oil droplet is rather immobile, and the moving particle takes an off-center position $\qb$. 

In the frame with the origin at the droplet center, the equilibrium shape is given by 
\ba
  u(\rb) &=& u_0 \ln\frac{R}{|\rb-\qb|} + u_2 \frac{R^2-r^2}{2R^2}  \nonumber\\
                     && -  u_0   \frac{\qb\cdot \rb}{R^2}  
                     -  u_0  \frac{2(\qb\cdot \rb)^2 - q^2r^2}{2R^4} +...,
\ea
where the additional terms assure the boundary condition $u(\Rb)=0$. In the following, they are truncated at the quadratic order in $q$.  

To simplify the calculation of the interfacial energy, we adopt the particle-fixed frame with coordinates $\rb'=\rb-\qb$, resulting in 
\ba
  u'(\rb') &=& u_0 \ln\frac{R}{r'} + u_2 \frac{R^2-r'^2}{2R^2}  
                     +  (u_0 + u_2)  \frac{\qb\cdot \rb'}{R^2}  \nonumber\\
                  &-&  u_0 \frac{q^2}{2R^2} 
                        - (2u_0 + u_2)\frac{2(\qb\cdot \rb')^2 - q^2r'^2}{2R^4} .
\ea
One readily verifies the boundary condition
\be
   u'(\Rb-\qb) = 0 ,
\ee
with $\Rb = R\eb_{r'}$. 

Defining the polar angle $\varphi$ through $\rb'\cdot\qb=r'q\cos\varphi$, and retaining the leading term in $q/R$ only, the  surface energy reads as
\be
    E[u'] = \frac{\gamma}{2} \int dS(\nabla u')^2 
       = E_0 +  \frac{k}{2}  q^2 ,
\ee
with the spring constant 
\be
  k =  4\pi \gamma  \frac{(u_0+u_2)^2}{R^2} =  4\pi \gamma  \tan^2\theta_\mathrm{C}. 
  \label{eq:spring}
\ee

% \section{Particle propulsion speed in thin film?}
% \textcolor{red}{optional if we write $v_p$ with prefactor using $u_\mathrm{w}, a, b$ }
% \begin{equation}
%     v_p=...?
% \end{equation}

\section{Speed and laser illumination intensity}\label{sec:speedandpower}
\begin{figure}[h]
\includegraphics[width=8cm]{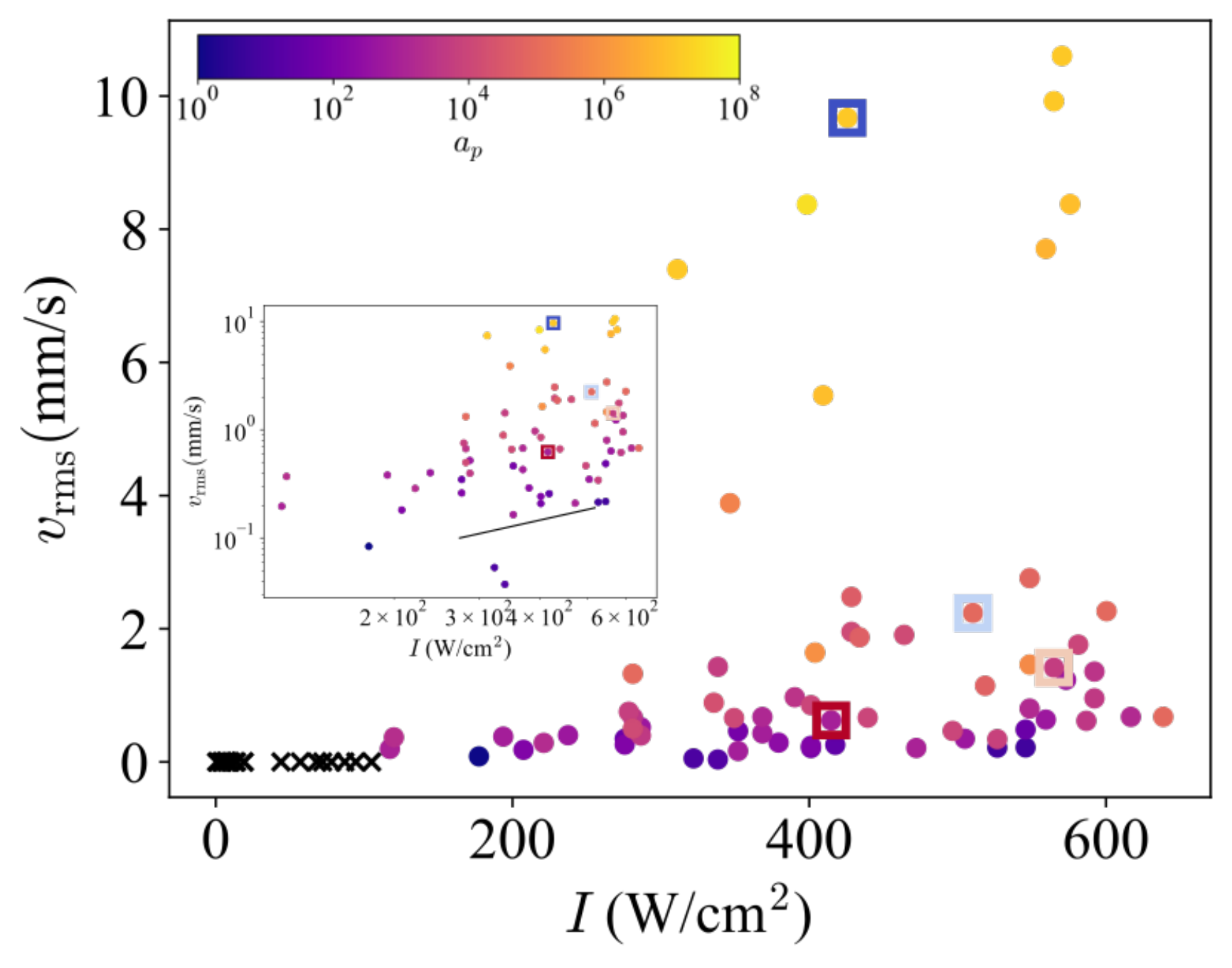}
\caption{\label{fig:vrms-power} (a) Laser illumination intensity and typical speed $v_\mathrm{rms}$. No significant motion was observed for ``$\times$.'' The red and blue squares correspond to Fig.~2 and 3(a) in the main article. Inset is in log scale with the line representing a guide to the eyes with a slope of 1. The color code refers to the spectral amplitude ${a_p}$.}
%(b) Laser illumination intensity and the maximum speed. (c) Mean radial lengths $\bar{r}$ and droplet radii $R$.}

\end{figure}

The thermal Marangoni-driven JPs at the interface move at a constant speed proportional to the laser illumination intensity $v_p \propto I$ both at a planar interface~\cite{dietrich_microscale_2020,wurger_thermally_2014} and at curved interfaces~\cite{ben_zion_cooperation_2022, ganesh_dynamics_2023}. As discussed in the main text, our case shows a time-dependent speed by the coupling with the capillarity. Still, we can find a correlation between the speed and the laser intensity, as shown in Fig.~\ref{fig:vrms-power}. Under a small illumination intensity, the JP remained static. Because the threshold for the motion might depend on the confined geometry of droplets, there is a coexistence region.
%We see a correlation but between intensity and $v_\mathrm{rms}$ which is not obvious but better than $v_\mathrm{max}$: the maximum speed for each sample. As we explained in the main text, $\bar{r}$ is almost constant because of the limited laser illuminating area, but with a little positive correlation with $R$, as shown in Fig.~\ref{fig:vrms-power}.

\section{Signal processing for the power spectral density (PSD)}
To extract the peak frequency $f_p$ in the main text, first, we smoothed the trajectory and got the velocity. To remove noise in a low-frequency region, we computed the PSDs of velocity (Fig.~3(a)) by segmenting every two seconds, windowed by a Hanning window, and then averaged. 

\begin{figure}[h]
\includegraphics[width=8cm]{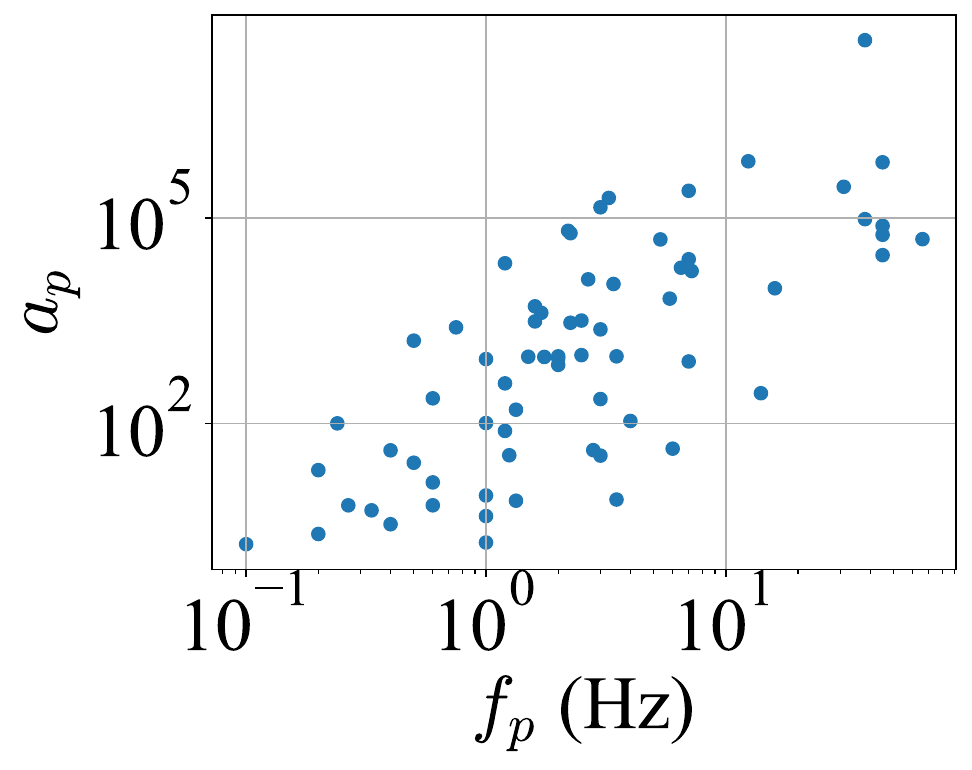}
\caption{\label{fig:fp-ap} Relationship between the relative amplitude of the PSD $a_p$ and the peak frequency $f_p$ of PSD.}
\end{figure}

To characterize various behaviors of JPs, we use $f_p$ as we explained in the main text, but the relative amplitude of the spectra $a_p:=S_{vv}(f)/S_{vv}(0)$ might also be used. As shown in Fig.~\ref{fig:fp-ap}, $a_p \propto f_p^2$. 

\section{Supplemental experiments on the relaxation processes}
\begin{figure}[h]
\includegraphics[width=\textwidth]{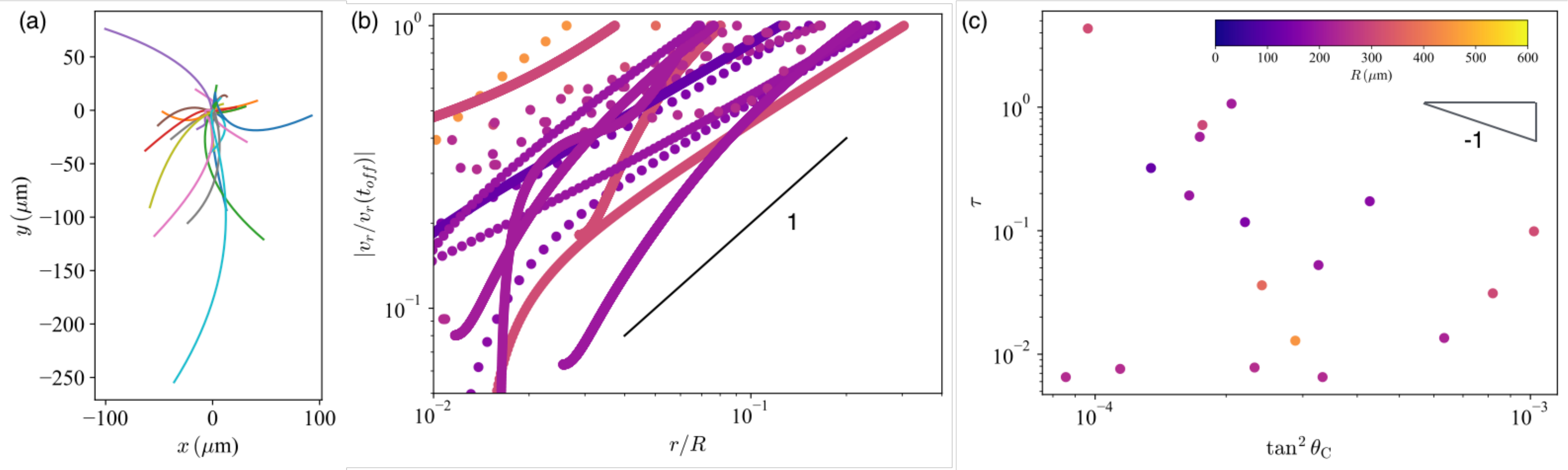}
\caption{\label{fig:S2} Relaxation process from active state. (a) Trajectories  (b) Radial speed as a function of radial length $r$, which might suggest that the force $\propto r^{\alpha}$. (c) The relaxation time by an exponential fit (using the first 3/4 of the time series. (b,c) is colored by the droplet radii $R$.}
\end{figure}

To clarify the centripetal force exerted on the JP due to the interfacial profile, we conducted relaxation experiments. We observed the motion of an off-center particle after switching off the laser at $t=t_\mathrm{off}$ after activation. Experimentally, the JP moved towards the center of the droplet, where the capillary force balances. The trajectories were not straight lines, as shown in Fig.~\ref{fig:S2}(a), which can be due to the inertia and the capillary force. Figure~\ref{fig:S2}(b) shows the temporal changes in radial length. They basically decay exponentially as for an overdamped harmonic oscillator, but some trajectories show oscillations around the equilibrium position, which means they are slightly underdamped. 

%(b) Radial position $r(t)$ normalized by the initial position.

The inertia is neglected in the relaxation process, then the viscous force balances the centripetal force. Because the viscous force is proportional to $v$, the velocity gives information about the capillary force. Here, we neglect the azimuthal component, and the normalized force is $|v_r/v_r(t_\mathrm{off})|\sim r$ as shown in Fig.~\ref{fig:S2}(c), suggesting a force by a harmonic oscillator.

Therefore, assuming the overdamped harmonic oscillator
\begin{equation}\label{eq:HO}
    k{\bf r }+\xi \dot{\bf r }=0,
\end{equation}
we estimated the relaxation time $\tau=\xi/k$ from the radial speed of the relaxation process. The timescale of $\tau$ ranges from $10^{-2}$\SI{}{\second}--$10^{0}$\SI{}{\second}, as shown in Fig.~\ref{fig:S2}(d). Remark that $\tau$ has little correlation with $R$ but a correlation with the contact angle $\theta_\mathrm{C}$. This is understood by the spring constant depending on $\theta_\mathrm{C}$, as seen in Eq.~\eqref{eq:spring}.
A tendency exists for strong confinement (large $\tan^2\theta_\mathrm{C}$) that showed more regular motions to result in a fast relaxation.

The relaxation time $\tau$ can be estimated via the Eqs.~\eqref{eq:HO} and \eqref{eq:spring} assuming $\xi=6\pi a\eta$, leading to
$\tau\sim 10^{-3}$\SI{}{\second}--$10^{-2}$\SI{}{\second} using physical values for oil ($\eta_{o}$ and $\gamma_{o}$). The reasons for the relaxation time scale discrepancy might be due to the higher-order deformations. More regular cases have less discrepancy, which suggests the profile is closer to the lower-order deformation.

\section{Supplemental experiments of Janus particles in thin film experiments}\label{sec:thinfilm}
To examine under which conditions the coupling of interfacial deformation and JP motion can occur, we conducted experiments with JPs in a thin flat film ($\sim\SI{10}{\micro\metre}=2a$) and thick flat film ($\sim\SI{60}{\micro\metre}=12a$), where $a$ is the radius of JPs.

We put a known volume of the oil suspension (\SI{20}{\micro\litre} or \SI{120}{\micro\litre}) containing JPs on distilled water in a glass Petri dish. We conducted experiments only at the center of the dish, where the effect of the meniscus is weak. We did not see the fringes without activation in the region of interest. When we began an illumination, the JP was placed at the illumination center, and tracked until it went out from the illuminated area. It should be noted that these experiments provide only very short observation times.

\begin{figure}[h]
\includegraphics[width=8cm]{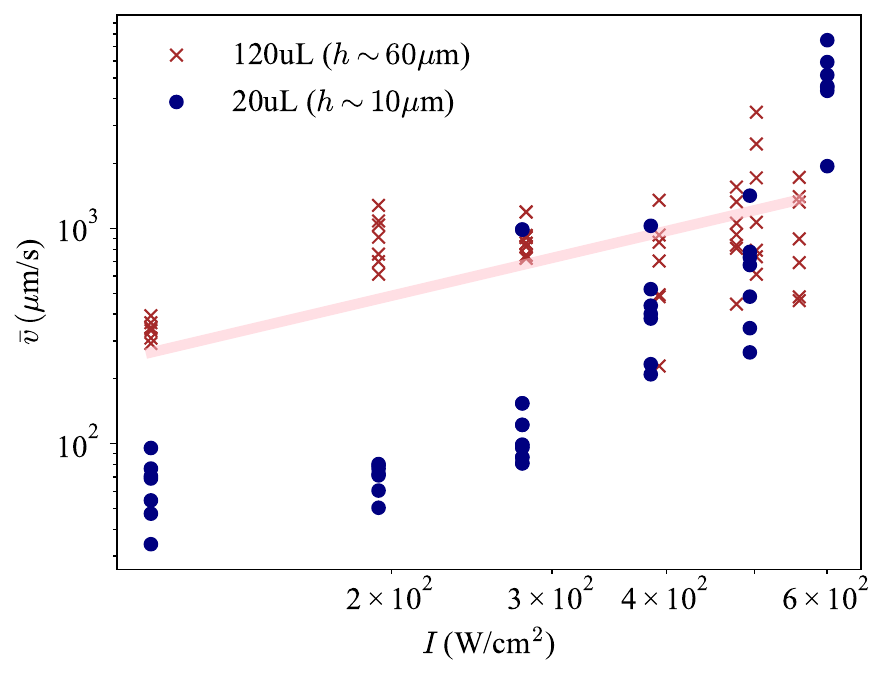}
\caption{\label{fig:thinfilm} Propulsion velocity $\bar{v}$ versus laser intensity $I$ . The pink line is the fitting to $\bar{v}\propto I$ for the thick film ($h\sim 60\SI{}{\micro\metre}$).}
\end{figure}

We measured the average speeds $\bar{v}$ for different laser illumination intensities $I$. In the thick film, the propulsion speed is almost $\bar{v}\propto I$ as shown in Fig.~\ref{fig:thinfilm}, and as expected for Marangoni surfers powered by single interfacial flow~\cite{dietrich_microscale_2020,wurger_thermally_2014,ganesh_dynamics_2023,ben_zion_cooperation_2022}. Because the JP is not trapped at the interface and might stay at an arbitrary vertical position in the oil film thickness \SI{60}{\micro\metre}, the speed varied from run to run. 

In contrast, in the thin film, the propulsion speed $\bar{v}$ is clearly different from $\bar{v}\propto I$ as shown in Fig.~\ref{fig:thinfilm}. Furthermore, we observed fringe propagation under the activation at all values of laser intensity in the ranges of $I=\qtyrange{109}{546}{\watt\per{\centi\metre}^2}$, while we did not observe fringes in the above thick film experiments. Unlike in a thick film, the thickness profile deforms by the activation in the thin film.

\section{Estimation of the capillary force and torque in steady circular motion}
\subsection{Estimation method in detail}
\begin{figure}[h]
\includegraphics[width=17cm]{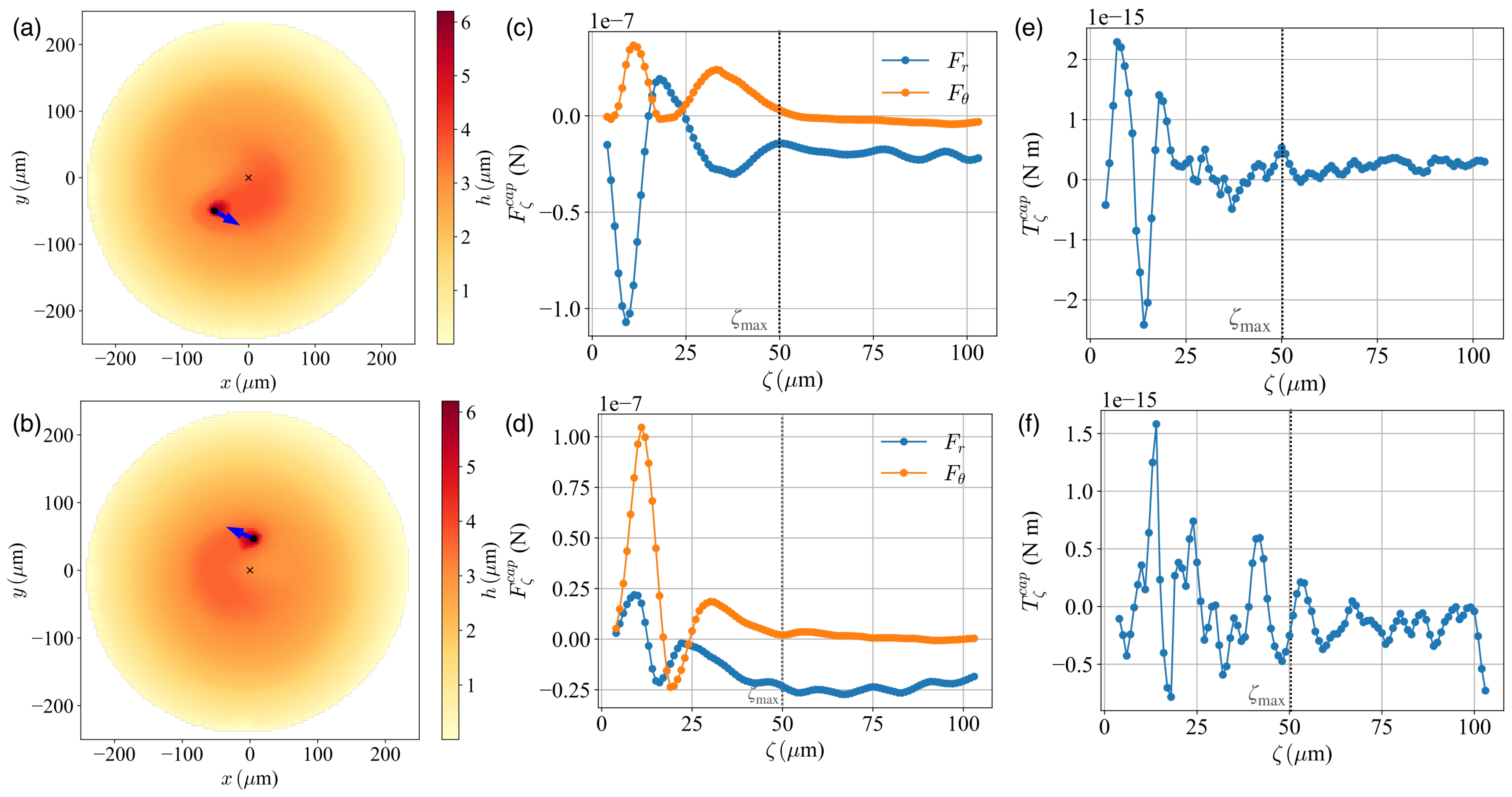}
\caption{\label{fig:capscenario}(a, b) The reconstructed height profile for two different moments of the same counterclockwise circular motion. ((a) is same as Fig.~4(a) in the main paper.) (c, d) Capillary forces around the circle $C_\zeta$ for (a, b) respectively. (e, f) Capillary torque around the circle $C_\zeta$ for (a, b) respectively.  }
\end{figure}

We take an example of periodic circular motion and estimate the capillary force and torque exerted on the JP solely from the interfacial profile. For this, we took contours from the fringes in a snapshot and assigned the height from the outer fringes. We assume that the most outward fringe corresponds to the phase difference $2\pi$, and the thickness is continuous and monotonous. Also, we assign $h=0$ at the edge of the droplet (three-phase contact line) and $h=2u_\mathrm{w}$ at the distance $r=r_\mathrm{w}$ from the JP position. For the thickness plot, we used a grid and linear interpolation. The reconstructed thickness profiles are shown in Fig.~\ref{fig:capscenario}(a, b). Using the expression Eq.~(3) in the main text, we can obtain a force field around the JP. We took the force application circle $C_\zeta$, and considered the net force and torque as a function of $\zeta$, as explained in the main text. The results for the two snapshots are shown in Fig.~\ref{fig:capscenario}(c, d) for force and (e, f) for torque, respectively. The detailed functional shape of ${\bf F}^{cap}_{\zeta}(\zeta)$ depends on the snapshot used, but the force and torque vanish for large $\zeta$. This means that the interfacial deformation within a certain distance from the JP contributes to the capillary force and torque exerted on the JP. Therefore, we set the cutoff distance $\zeta_\mathrm{max}=\SI{50}{\micro\metre}$ (The dashed lines in Fig.~\ref{fig:capscenario}(c--f)) for averaging to avoid spurious contributions.

\subsection{Estimated capillary force and torque}
The estimation results of the capillary force ${\bf F}^{cap}$ and torque $T^{cap}$, as well as the calculated viscous force $F_v$ and torque $T^{vis}$ are summarized in Table~\ref{tab:estimation}.
The JP feels a centripetal force $ F^{cap}_r<0$, which is comparable to the viscous force in bulk ($v_p\sim 1\SI{}{\milli\metre\per\second}$--$10\SI{}{\milli\metre\per\second}$ is used). 
The positive azimuthal force $F_\theta^{cap}>0$ means the JP is pushed forward. We often see a thinning at the tail of the JP, as seen also in Fig.~2(a,b), so the rear profile becomes steeper, which pushes the JP forward. The positive torque $T^{cap} > 0$ for the counterclockwise circular motion means an inward rotation of the JP polarity. The estimated value of the torque is comparable to the viscous torque ($\Omega=45$Hz rotation is used by assuming the JP is rotating at the same frequency as that of the circular motion). 
Note that the values remain similar in other snapshots of the dynamics, including their signs
\begin{table}[b]%The best place to locate the table environment is directly after its first reference in text
\caption{\label{tab:estimation}%
Force and torque estimation
}
\begin{ruledtabular}
\begin{tabular}{lccccc}
&
$F_r^{cap}$&
$F_\theta^{cap}$&
$T^{cap}$&
$F_v$&
$T^{vis}$
\\
\colrule
obtained via&
\multicolumn{3}{c}{the reconstructed thickness profile}&$6\pi a \eta_o v_p$ &$8\pi a^3 \eta_o \Omega$
\\
\colrule
estimation&$-3\cdot10^{1}\SI{}{\nano\newton}$ & $1\cdot10^{1}\SI{}{\nano\newton}$ & $3 \cdot 10^{-16}\SI{}{\newton \metre}$ & $ 1\SI{}{\nano\newton}$--$10\SI{}{\nano\newton}$& $ 4 \times 10^{-16}\SI{}{\newton \metre}$\\
\end{tabular}
\end{ruledtabular}
\end{table}

\section{Supplemental experiments using symmetric particles}\label{sec:coreshell}
    To clarify the role of polarity in the motion of Janus particles, we used core-shell particles of the same size as Janus particles (\SI{10}{\micro\meter}), made of polystyrene spheres fully coated with gold. (SG10u, nanocs; coating thickness: \qtyrange{5}{10}{\nano\meter}).

For droplets with single core-shell particles, we confirmed the fringe changes by laser illumination, due to heating which changes the profile close to the particle (see also discussion in Sec.\ref{sec:dewet}). Importantly, {\bf no motion was observed even under high-intensity illumination over $\SI{550}{\watt\per({\centi\metre}^2)}$}. Although the gold thickness is around ten times smaller than our JP's, a cluster of core-shell particles can lead to a self-rotation or a regular motion like a JP. Therefore, we believe a significant asymmetry is necessary for self-propulsion, which is different from the instability of global flow by a heated particle~\cite{Koleski2020}.

\section{Interfacial profile changes by heating}\label{sec:dewet}
%\textcolor{red}{maybe word ``dewetting'' can be not appropriate. we never know the profile close to the JP... write a possible scenario as discussion  like flattening around the JP, dewetting ~\cite{fiegel_wetting_2005},...}
The interfacial profile changes can be triggered by heating of the JP, as well as a hydrodynamic flow. In this section, we discuss that a heating effect exists realistically. The temperature gradient changes both the interfacial tension spatially $\gamma(r)$ as well as the boundary condition of wetting angle $\theta_\mathrm{W}$ which is determined by the interfacial tensions $\gamma$ for the JP in contact with the fluids (air, oil, and aqueous phase). 

%the dewetting can be a probable scenario for the following reasons. First, we observed the complete dewetting of the JP from a thin droplet experimentally (See the time-lapse image in ?). Second, 
As shown in the following, we observed an inward fringe displacement for a symmetric heated particle in a droplet. In addition, we observed the outward propagation of fringes around a JP by a deformation of a thin flat film under a small activation. Both suggest an interfacial profile changed by heating. 

We should remark that the values $\theta_\mathrm{W}$ and $\Delta\theta_\mathrm{W}:=\theta_\mathrm{W}(T+\Delta T)-\theta_\mathrm{W}(T)$, as well as the detailed profile close to the JP, are not accessible so that we do not know whether the interfacial profile changes are caused by wetting $\Delta\theta_\mathrm{W}<0$ or dewetting $\Delta\theta_\mathrm{W}>0$.

\subsubsection{The inward fringe displacement for a symmetric heated particle in a droplet}
\begin{figure}[h]
\includegraphics[width=7cm]{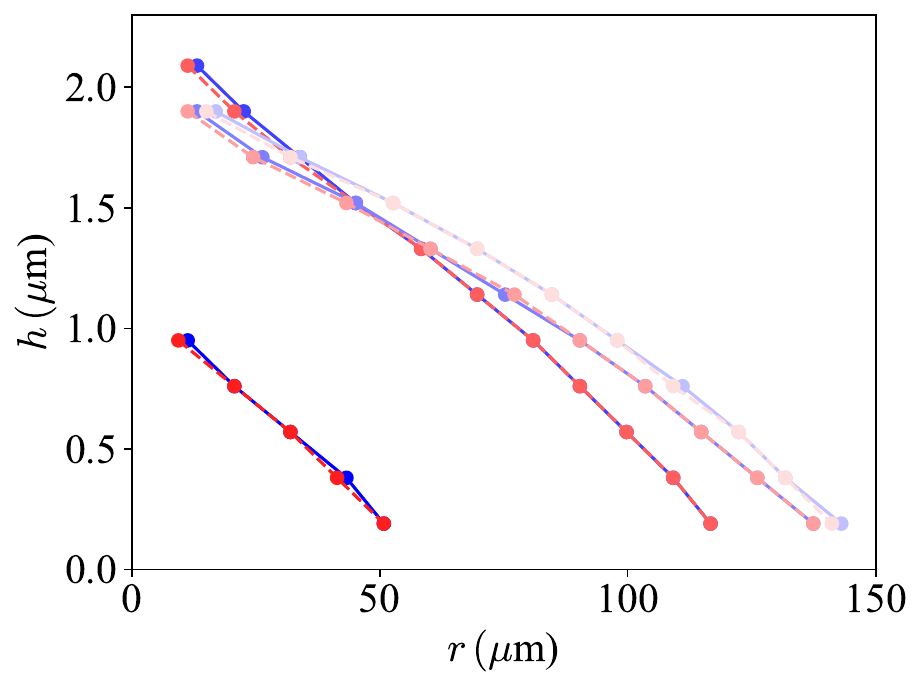}
\caption{\label{fig:coreshell} Fringe symmetric displacement by heated coreshell particle. Blue: passive, Red: active (from the darker line and symbol, $I=382, 410, 483, 614\SI{}{\watt\per{\centi\metre}^2}$). The symbols correspond to the positions of fringes. The fringes closest to the JP in our resolution moved around \SI{2}{\micro\metre}.}
\end{figure}

First, we observed the inward fringe displacement for a symmetric particle in a droplet (See the Method for Sec. \ref{sec:coreshell}). As shown in Fig.~\ref{fig:coreshell}, the fringe close to the particle moved inward in all cases. Note that because the thickness of the gold layer is 10-time thinner compared with that of our JPs, the heating effect is not significant. Still, we observed an inward fringe displacement of around $\sim\SI{2}{\micro\metre}$ for the fringe nearest to the particle. 

The fringe inward displacement is an interfacial deformation, and is not due to a change in the refractive index. We can exclude the possibility of the refractive index changes in the following.

The refractive index $n$ usually decreases with temperature.
For alkanes, the refractive index depends on density: $n=5\cdot 10^{-4}\cdot\rho+1.05$~\cite{Winoto2014}, where $\rho$ is in [$\SI{}{\kilo\gram\per \metre^3}$].  The density $\rho$ depends on temperature: $\rho=-0.648T_{deg}+952.50$ around $T_{deg}=20\SI{}{\degreeCelsius}$--$40\SI{}{\degreeCelsius}$~\cite{Santos2017}.
Therefore, $n_{\it o}=-3.24\cdot 10^{-4}T_{deg}+1.53$.
Thus, the refractive index of tetradecane $n_{\it o}$ can be 1.428, 1.427, 1.425 at 30, 35, 40\SI{}{\degreeCelsius}.

We observed an inward fringe motion of ($\sim\SI{2}{\micro\metre}$) in the droplet with a coreshell particle. To judge if this is due to the interfacial shape change or the refractive index change by a temperature rise, we consider the following situation:

Let us suppose there is no shape change but only a refractive index change to $n_{\it o}'$=1.425 at 40\SI{}{\degreeCelsius} from 20\SI{}{\degreeCelsius}. The $m$th fringe height changes by $\Delta h_m=\frac{m\lambda}{2}(\frac{1}{n_{\it o}'}-\frac{1}{n_{\it o}})$.
Locally, a small height difference $\Delta h$ obeys $\Delta h = \frac{\lambda \Delta r}{2n_{\it o}\Delta r_0}$, where $\Delta r_0$ and $\Delta r$ are the typical fringe interval and the fringe displacement. We can take $\Delta r_0\sim$\SI{10}{\micro\metre} in a large case and consider the above fringe height change $\Delta h_m$, then the $m$th fringe displacement is $\Delta r \sim m\cdot$(\SI{0.03}{\micro\metre}). In an example, $m=11$th fringe (the most inward fringe loop detectable) moved \SI{2}{\micro\metre}, which is much larger than the value expected via refractive index change: \SI{0.33}{\micro\metre}.
Therefore, the fringe inward motion in the droplet with a core-shell particle is not due to the refractive index changes but a real interfacial deformation. 

\subsubsection{The outward propagation of fringes around a JP in a thin flat film}
In thin film experiments in Sec.~\ref{sec:thinfilm}, the fringes propagated outward away from the center of the particle. The area with visible fringes increased and then started decreasing when the JP reached the edge of the illumination area.
At a small intensity ($I=109\SI{}{\watt\per{\centi\metre}^2}$), the fringe propagation started prior to the motion. In addition, the interfacial deformation by an active JP particle might not be symmetric. Therefore, it is plausible to consider this fringe propagation as the result of the heating effect. The fringe propagation speed was $v_\mathrm{fringe}\sim\SI{1}{\milli\metre\per\second}$ at $I=109\SI{}{\watt\per{\centi\metre}^2}$. Let us remark that this speed is much slower than the thermal capillary wave of thin films~\cite{zhao_laser-induced_2024} ($\mathcal{O}(\SI{1}{\metre\per\second})$), which is not our case. The interfacial profile changes by heating can explain the fringe outward motion.

\section{Non-steady behavior with a large dilation}
The importance of droplet confinement can be gauged via experiments with non-steady behaviors, which were observed only for a tiny droplet under a large illumination intensity. 
\begin{figure}[h]
\includegraphics[width=17cm]{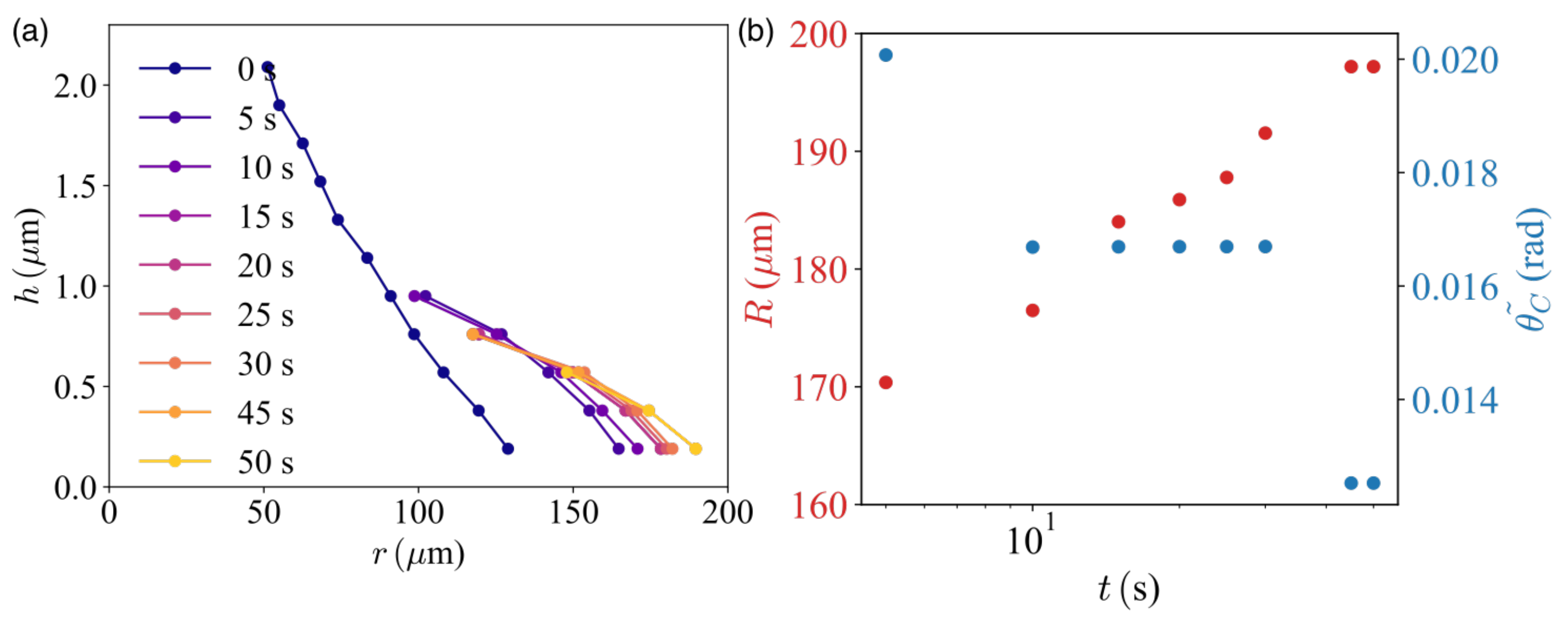}
\caption{\label{fig:dilation} Non-steady behavior under the illumination at $I=380\SI{}{\watt\per{\centi\metre}^2}$. (a) The thickness profile changed with time. (b) Droplet radius $R$ increases, and the contact angle decreases.}
\end{figure}

In a Supplementary Video 3 (see  Fig.~\ref{fig:dilation}(a,~b)), the droplet radius $R(t)$ increased 46\% and the approximated contact angle $\tilde{\theta}_\mathrm{C}:=\arctan(-\mathrm{d}h/\mathrm{d}r\vert_{r=R}) (\sim \theta_\mathrm{C})$ decreased by 38\% in \SI{50}{\second}. This dilation changed the confinement, leading to a transition from a regular regime to an irregular one. We note that the final radius of the droplet did not change after switching off the activity. This means that the droplet goes from a metastable state to another metastable state at the same temperature. We believe a viscous force drove the dilation due to the JP motion. The force per unit length for the droplet edge motion $\gamma |\cos\tilde{\theta}_\mathrm{C}^{final}-\cos\tilde{\theta}_\mathrm{C}^{initial}|$ is the order of $10^{-6}\SI{}{\newton\per\metre}$ for this large dilation case. In contrast, $\Delta T=\SI{1}{\kelvin}$ heating will give $|\mathrm{d}\gamma/\mathrm{d}T\cdot \Delta T|$, resulting in the force in the order of $10^{-4}\SI{}{\newton\per\metre}$. The viscous force $\sim \eta v$ gives $10^{-6}\SI{}{\newton\per\metre}$ for $v=\SI{1}{\milli\metre\per\second}$ whose value corresponds to the one for dilation. In addition, the rise in temperature due to the JP would not happen globally with the existence of the aqueous subphase working as a heat reservoir. %The hypothetical temperature rise in steady state $\Delta T \rightarrow T+\Delta T$ leads to each interfacial tension change in the Young equation Eq.~\ref{eq:Young} by $\gamma=\mathrm{d}\gamma/\mathrm{d}T\cdot\Delta T$, but the small temperature rise such as $\Delta T=\SI{0.6}{\kelvin}$ lead to the breakdown of the condition for the Young equations Eq.~\ref{eq:Young} of lens-like droplets, resulting in the formation of oil thin film in equilibrium. Therefore, droplet dilation in two-dimensions was not driven by the global temperature rise but is due to the motion of the active JP.

\section{Transient trajectories of the model}
\begin{figure}[h]
\includegraphics[width=8cm]{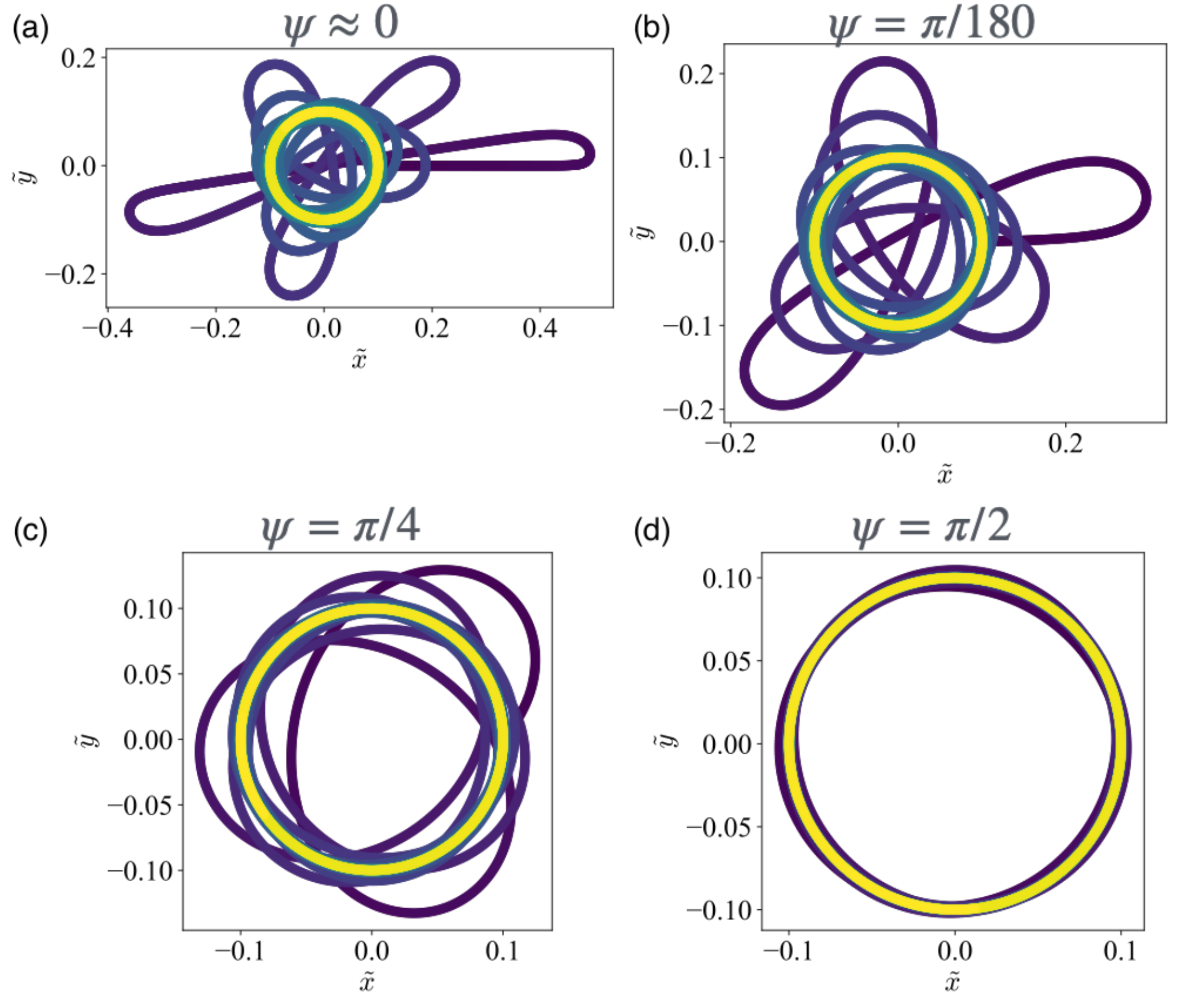}
\caption{\label{fig:model_nonsteady} The transient trajectories suggest one possible scenario for the back-and-forth motion. $A=100,\, (\tilde{x}(0),\, \tilde{y}(0))=(0.1,0)$. Non-steady trajectories depend on how much the initial orientation $\hat{\bf p}(0)=(\cos\psi,\sin\psi)$ has an azimuthal component. The initial polarity angles $\psi$ are changed. All trajectories go to the same limit cycle (yellow).}
\end{figure}

In the model described in the main text, the particle relaxes to a limit cycle for sufficiently large $A$. When the initial velocity is closer to the radial direction, the effect of torque to rotate the particle is smaller, which takes a longer time to relax to the periodic state. In that process, back-and-forth motion-like trajectories can be seen as shown in Fig.~\ref{fig:model_nonsteady}. 

Nonetheless, an alternative scenario could involve out-of-plane rotation ($\bf{\Omega}$ parallel to $xy$ plane), which goes beyond the scope of the aforementioned model.

\section{Supplemental experiments adding tracers}
To obtain information on the fluid flow by an activation, \SI{0.6}{\micro\meter} TiO$_2$ tracer particles were added to the oil phase. As general problems, we cannot conduct particle tracking velocimetry with enough traces; tracers tend to stick on the surface of the Janus particle by capillary attraction, which changes the Janus particle motion itself. Thus, we put a small amount of tracer particles, and focused only on free tracers.

\begin{figure}[h]
\includegraphics[width=0.75\textwidth]{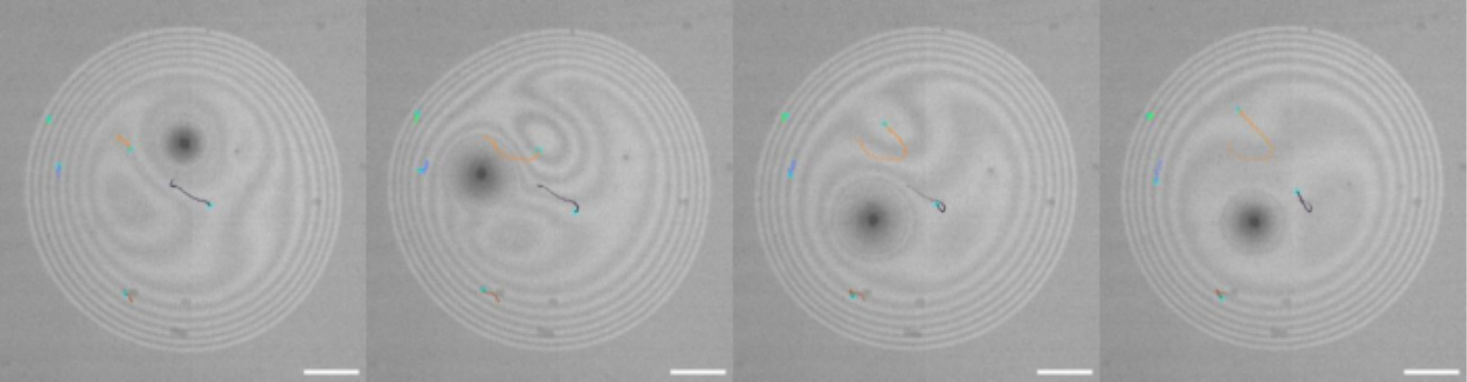}
\caption{\label{fig:tracer} Montage every $\SI{0.25}{\second}$ ($I=344\SI{}{\watt\per{\centi\metre}^2}$). Trajectories of some tracers are marked. The orange-marked trace went into the fringe loop that JP made at its previous position. Scalebar=\SI{100}{\micro\metre}}.
\end{figure}

We observed that only tracers close to the active JP moved, which suggests the flow is generated locally adjacent to the JP. Moreover, a tracer went to the center of a convex or concave structure (Fig.~\ref{fig:tracer}, as well as Supplementary Video 4.): nearly concentric fringe loops that contained no JP when the structure was vanishing. It suggests that the fringe loops mean a concave structure which get slowly filled with liquid. 
% \section{other memo}
% Reynolds number for our system at maximum: $\mathrm{Re}=(0.762\SI{}{\gram\per\milli\litre}\cdot \SI{10}{\milli\metre\per\second}\cdot \SI{10}{\micro\metre})/(\SI{2.81}{\milli \pascal \second}) \approx 0.027$.

\section{List of Supplemental Videos}
\begin{itemize}
    \item \url{S1_regular.mov} Regular (periodic, circular) motion, $I=426\SI{}{\watt\per{\centi\metre}^2}$.
    \item \url{S2_irregular.mov} Irregular motion,$I=415\SI{}{\watt\per{\centi\metre}^2}$.
     \item \url{S3_largedeformation.mov} Large deformation of the droplet.\\ Tracers (TiO$_2$, $\SI{0.6}{\micro\metre}$) were in the oil phase. The acceleration of the JP led to the formation of concentric fringe loops, which did not contain the JP inside. These fringe loops are a concave structure, indicating thinning at the rear side of the JP. This modulation in thickness, which occurs away from the JP, is likely driven primarily by hydrodynamic flow. $I=344\SI{}{\watt\per{\centi\metre}^2}$. x0.25 play.
    \item \url{S4_escape.mov} JP escaping from the oil phase in an extremely thin case, $I=647\SI{}{\watt\per{\centi\metre}^2}$.
    \item \url{S5_nonsteady.mov} Non-steady case with a large dilation, $I=380\SI{}{\watt\per{\centi\metre}^2}$.
\end{itemize}

\section{Material data}

\begin{table}[h]
 \caption{Values at 20\SI{}{\degreeCelsius}}
 \label{table:materialliquid}
 \centering
  \begin{tabular}{clllll}
   \hline
      & $\gamma\SI{}{[\milli\newton\per\metre]}$ &  $\rho\SI{}{[g/mL]}$ & n & $\eta [\SI{}{\milli \pascal \cdot\second}]$ & $\kappa [\SI{}{\watt\per (\metre \cdot\kelvin)}]$  \\
   \hline \hline
   tetradecane & 26.53  & 0.762 & 1.429 &2.81 &0.14 (*the value of decane)\\
   water & 72.8 &  1 & 1.33 &1 & 0.61 \\
   \hline
  \end{tabular}
\end{table}

\begin{table}[h]
 \caption{Values related to interfaces}
 \label{table:materialint}
 \centering
  \begin{tabular}{lll}
   \hline
      $\gamma_\mathrm{\it o}\SI{}{[\milli\newton\per\metre]}$ & $\gamma_\mathrm{\it w}\SI{}{[\milli\newton\per\metre]}$ & $\gamma_\mathrm{\it ow}\SI{}{[\milli\newton\per\metre]}$ \\
    \hline
    26.53 & 72.8 & 46.28\footnote{calculated using the Young relation and the measured contact angles}\\
   \hline \hline
   $\mathrm{d}\gamma_\mathrm{\it o}/\mathrm{d}T$ \SI{}{[\milli\newton\per(\metre\cdot\kelvin)]} &$\mathrm{d}\gamma_\mathrm{\it w}/\mathrm{d}T$\SI{}{[\milli\newton\per(\metre\cdot\kelvin)]}
   &$\mathrm{d}\gamma_\mathrm{\it ow}/\mathrm{d}T$\SI{}{[\milli\newton\per(\metre\cdot\kelvin)]}\\
   \hline
   -0.144\cite{queimada_surface_2001}&-0.151&-0.08\cite{zeppieri_interfacial_2001}\\
   \hline
  \end{tabular}
\end{table}

\begin{table}[h]
 %\caption{value}
 \label{table:materialsolid}
 \centering
  \begin{tabular}{cccc}
   \hline
      &  $\rho\SI{}{[g/mL]}$ & n  & $\kappa[\SI{}{\watt\per (\metre \cdot \kelvin)}]$ \\
   \hline \hline
   Polystyrene & 1 & 1.58 &1.0 \\
   Cr &7.19 &-&92\\
   Au &19.3 &-&315\\
   Our JP &1.56 (effective)& - & -\\
   \hline
  \end{tabular}
\end{table}

%\bibliographystyle{apsrev4-2}
%\bibliography{main.bib}% Produces the bibliography via BibTeX.